\documentclass[twocolumn]{aa} % for a paper on 2 column
\def\ms{\hbox{m$\;$s$^{-1}$}}

\usepackage{natbib}
\usepackage{graphicx}
\begin{document}
\title{MHD wave propagation from the sub-photosphere to the corona in an arcade-shaped magnetic field with a null point}

\titlerunning{Wave propagation from the sub-photosphere to the corona}

\author{I.C. Santamaria\inst{1,2}, E. Khomenko\inst{1,2,3}, M. Collados\inst{1,2}}
\authorrunning{I.C. Santamaria et al.}

\institute{Instituto de Astrof\'{\i}sica de Canarias, 38205 La Laguna, Tenerife, Spain
\and Departamento de Astrof\'{\i}sica, Universidad de La Laguna, 38205, La Laguna, Tenerife, Spain 
\and Main Astronomical Observatory, NAS, 03680, Kyiv, Ukraine}

\date{Received; Accepted }

\abstract {}
{The aim of this work is to study the energy transport by means of MHD waves propagating in quiet Sun magnetic topology from layers below the surface to the corona. Upward propagating waves find obstacles, such as the equipartition layer with plasma $\beta=1$, the transition region and null points, and get transmitted, converted, reflected and refracted. Understanding the mechanisms by which MHD waves can reach the corona can give us information about the solar atmosphere and the magnetic structures.}
{We carry out two-dimensional numerical simulations of wave propagation in a magnetic field structure that consists of two vertical flux tubes with the same polarity separated by an arcade shaped magnetic field. This configuration contains a null point in the corona, that significantly modifies the behaviour of the waves as they pass near it. 
}
{We describe in detail the wave propagation through the atmosphere under different driving conditions. We also present the spatial distribution of the mean acoustic and magnetic energy fluxes for the cases in which these calculations are possible and the spatial distribution of the dominant frequencies in the whole domain.}
{We conclude that the energy reaches the corona preferably along almost vertical magnetic fields, that is, inside the vertical flux tubes. This energy has an acoustic nature.
Most of the magnetic energy keeps concentrated below the transition region due to the refraction of the magnetic waves and the continuous conversion of acoustic-like waves into fast magnetic waves in the equipartition layer located in the photosphere where plasma $\beta$ = 1. However, part of the magnetic energy reaches the low corona when propagating in the region where the arcades are located, but waves are sent back downwards to the lower atmosphere at the null point surroundings. This phenomenon, together with the reflection and refraction of waves in the TR and the lower turning point, act as a re-feeding of the atmosphere, that keeps oscillating during all the simulation time even if a driver with a single pulse was used as initial perturbation. In the frequency distribution, we find that high frequency waves can reach the corona outside the vertical flux tubes.}

\keywords{Sun: magnetic fields; Sun: oscillations; Sun: photosphere; Sun: chromosphere; Sun: MHD waves}

\maketitle

%________________________________________________________________
\section{Introduction}

Quiet sun magnetic structures are much less organized and have less intense magnetic field compared to sunspots \citep{SanchezAlmeida2011}. At the same time, quiet Sun covers at least about 90\% of the solar surface \citep{Sheeley1967, Lites+etal1996,Lin+Rimmele1999, Khomenko+etal2003_2,TrujilloBueno+etal2004} independently of the activity level, which makes quiet Sun magnetic structures to be important contributors to the solar magnetism. These weak structures are very dynamic, constantly perturbed by convection and propagating waves \citep[see][to name a few]{Muller+etal1994, Berger+Title1996, Krijer+etal2001, DePontieu+etal2003, Vecchio+etal2007, Vecchio2009,  Centeno+etal2009, FujTsun09, Kontogiannis2010a, Kontogiannis2010b, Kontogiannis2011, MartinezGonzalez2011, Chitta2012}. 

Waves are stochastically excited by turbulent convection at the base of the photosphere \citep{Goldreich+Keeley1977, Balmforth1992, Nordlund+Stein2001}. The magnetic field concentrations embedded in the intergranular lanes suffer foot point motions and this drives waves in these structures through the solar atmosphere \citep{Hasan+etal2000, Kato+etal2011}. Depending on the foot point motion, different waves are going to be excited. There is a large number of works on numerical simulations that reproduce these different scenaria: horizontal motions producing transverse kink waves; pressure fluctuations producing longitudinal waves; twisting motions generating torsional Alfv\'{e}n waves; observationally-driven or impulsive motions \citep{Ulmschneider+etal1991, Choudhur+etal1993,Rosenthal+etal2002, Bogdan+etal2003, Hasan+etal2003, Hasan+etal2005, DePontieu+etal2004, DePontieu+etal2005, Khomenko+etal2008,Vigeesh2009,Heegland+etal2009, Khomenko+Cally2011,Fedun+etal2011a, Fedun+etal2011b, Nutto+etal2012}. With some exceptions \citep{Choudhur+etal1993, DePontieu+etal2005, Fedun+etal2011b}, these works model the wave propagation in the photospheric and chromospheric layers. 

The dominant oscillations observed in the quiet Sun photosphere are acoustic-gravity waves with a period of 5-minute \citep{Leighton+etal1962, Ulrich1970}. As these waves propagate upwards the frequency distribution changes. Essentially 3-minute period waves are observed in the quite chromosphere inside super granular lanes \citep{Deubner+Fleck1990, Lou1995, Hoekzema+Rutten1998, Rutten+Uitenbroek1991, Lites1993},  and 5-minute oscillations are detected in and around magnetic elements in facular regions  \citep{DePontieu+etal2004, DePontieu+etal2005, Hansteen2006, Jefferies+etal2006, McIntosh+Jefferies2006, Centeno+etal2006, Kostik+Khomenko2013} and network regions \citep{Lites1993, Krijer+etal2001, DePontieu+etal2003, Bloomfield+etal2006, Tritschler2007, Vecchio+etal2007}, with some differences in behaviour, see recent review by \citet{Khomenko+CalvoSantamaria2013}. Both 3- and 5-minute oscillations appear in the corona  \citep{DeMoortel+Nakariakov2012}. The distribution of the dominant frequency of oscillations along the atmosphere appears to be different inside the quiet network cells and network borders and facular regions, containing stronger magnetic fields. There are observations of the 5-min waves channelled along the inclined magnetic field lines to the chromosphere in facular regions (via so-called ``ramp-effect'' or ``magnetoacoustic-portals''), which have been explained by the reduction of the acoustic cut-off frequency in inclined magnetic fields \citep{Michalitsanos1973, Suematsu1990, DePontieu+etal2005, Jefferies+etal2006, Heggland2007, Heggland2011}. \citet{Centeno+etal2009} and \citet{Khomenko+etal2008} proposed that radiative losses in thin flux tubes can also help propagating long-period oscillations vertically upward \citep[see][]{Roberts1983}. The latter mechanism, however, was not found to be dominant in the simulations by  \citet{Heggland2011}, with a complex treatment of radiative transfer. 

\parskip 0pt

As the waves propagate upwards they find obstacles, such as the equipartition layer where plasma $\beta=1$, the layer where the local acoustic cut-off frequency is equal to the wave frequency, the steep temperature gradient at the transition region, etc. In these regions waves suffer transformations, refraction, and reflection. Since the atmospheric properties change as we move from one layer to another, and even from one column to another, these effects can appear at different horizontal and vertical positions. For instance while in the photosphere the gas pressure dominates, the magnetic pressure becomes dominant from the middle chromosphere upwards. 

\parskip 0pt

Particularly interesting is the question of how, and in which amount, the energy of waves excited in sub-photospheric layers can reach the chromosphere and corona, and whether the wave modes reaching there can be dissipated to convert their energy into heat, i.e. how the solar atmospheric layers are magnetically and energetically connected by means of waves. Only few numerical studies perform simulations of waves in the whole domain from the photosphere to the chromosphere, transition region and corona. \citet{Fedun+etal2011b}, in their 2D simulations of high-frequency wave propagation, find that magneto-acoustic waves in a wide range of frequencies can effectively leak energy into the corona. These authors considered an isolated magnetic flux tube that becomes vertical in the chromosphere. In the present work we study  how the energy reaches the chromosphere and corona by exciting waves below the photosphere with different perturbations. We choose a complex magnetic field that consists of vertical flux tubes of the same polarity separated by an arcade-shaped magnetic field, resembling a network and internetwork regions, with wave periodicities in the 3-5 min regime. We discuss in detail the behaviour of different wave modes in such complex situation. The length of our simulated time series allows us to reach the stationary regime in two of the three considered cases and quantify the energy flux reaching the corona, as well as the frequency distribution of waves with height and horizontal distance.

%%%%%%%%%%%%%%%%%%%%%%%%%%%%%%%%%%%%%%%%%%%%%%%
\section{Numerical Method}
%%%%%%%%%%%%%%%%%%%%%%%%%%%%%%%%%%%%%%%%%%%%%%%

We solve the two-dimensional ideal magnetohydrodynamic equations of conservation of mass, momentum, energy, and the induction equation for the magnetic field: 

\begin{equation} \label{eq:den}
\frac{\partial\rho}{\partial t}+\nabla(\rho{\bf v})= \Big [ \frac{\partial\rho}{\partial t} \Big ]_{\rm diff} \,, 
\end{equation}
\begin{equation} \label{eq:mom}
\frac{\partial (\rho{\bf v})}{\partial t}+\nabla\Big [\rho{\bf vv}+\Big (p+\frac{{\bf B}^2}{2 \mu_0}\Big ){\bf I}-\frac{{\bf B}{\bf B}}{\mu_0}\Big ]=\rho{\bf g}  + \Big [ \frac{\partial (\rho{\bf v})}{\partial t} \Big ]_{\rm diff}    \,,
\end{equation}
\begin{equation} \label{eq:ei}
\frac{\partial p}{\partial t}+\textbf{v} \nabla p + \gamma p \nabla\textbf{v} = \Big [\frac{\partial p}{\partial t} \Big]_{\rm diff} \,,
\end{equation}
\begin{equation} \label{eq:ind}
\frac{\partial {\bf B}}{\partial t}=\nabla\times ({\bf v} \times {\bf B}) + \Big [ \frac{\partial {\bf B}}{\partial t}\Big ]_{\rm diff} \,,
\end{equation}

\noindent where all the notations are standard. We use an ideal equation of state and $\gamma = 5/3$. The terms with subscript ``diff'' are the artificial hyper-diffusive terms required for the code's numerical stability, see \citet{Felipe2010}. After removing the equilibrium condition (see below), we solve these non-linear equations by means of the code {\sc mancha} which is described in detail in \citet{Khomenko+Collados2006} and \citet{Felipe2010}. The numerical code solves the equations for perturbations. Through the present study we keep the amplitude of the perturbations small enough to concentrate on the linear regime. The non-linear effects will be reported in a forthcoming study.

%%%%%%%%%%%%%%%%%%%%%%%%%%%%%%%%%%%%%%%%%%%%%%%
\subsection{MHS model}
%%%%%%%%%%%%%%%%%%%%%%%%%%%%%%%%%%%%%%%%%%%%%%%

For simplicity, we have chosen a potential magnetic field structure. Such magnetic field does not interact with the equilibrium atmosphere. This allows us to compute the hydrostatic equilibrium model separately from the magneto-static equilibrium.

%%%%%%%%%%%%%%%%%%%%%%%%%%%%%%%%%%%%%%%%%%%%%%%
\subsubsection{Hydrostatic model}
%%%%%%%%%%%%%%%%%%%%%%%%%%%%%%%%%%%%%%%%%%%%%%%

\begin{figure*}
\centering
\includegraphics[width=12cm]{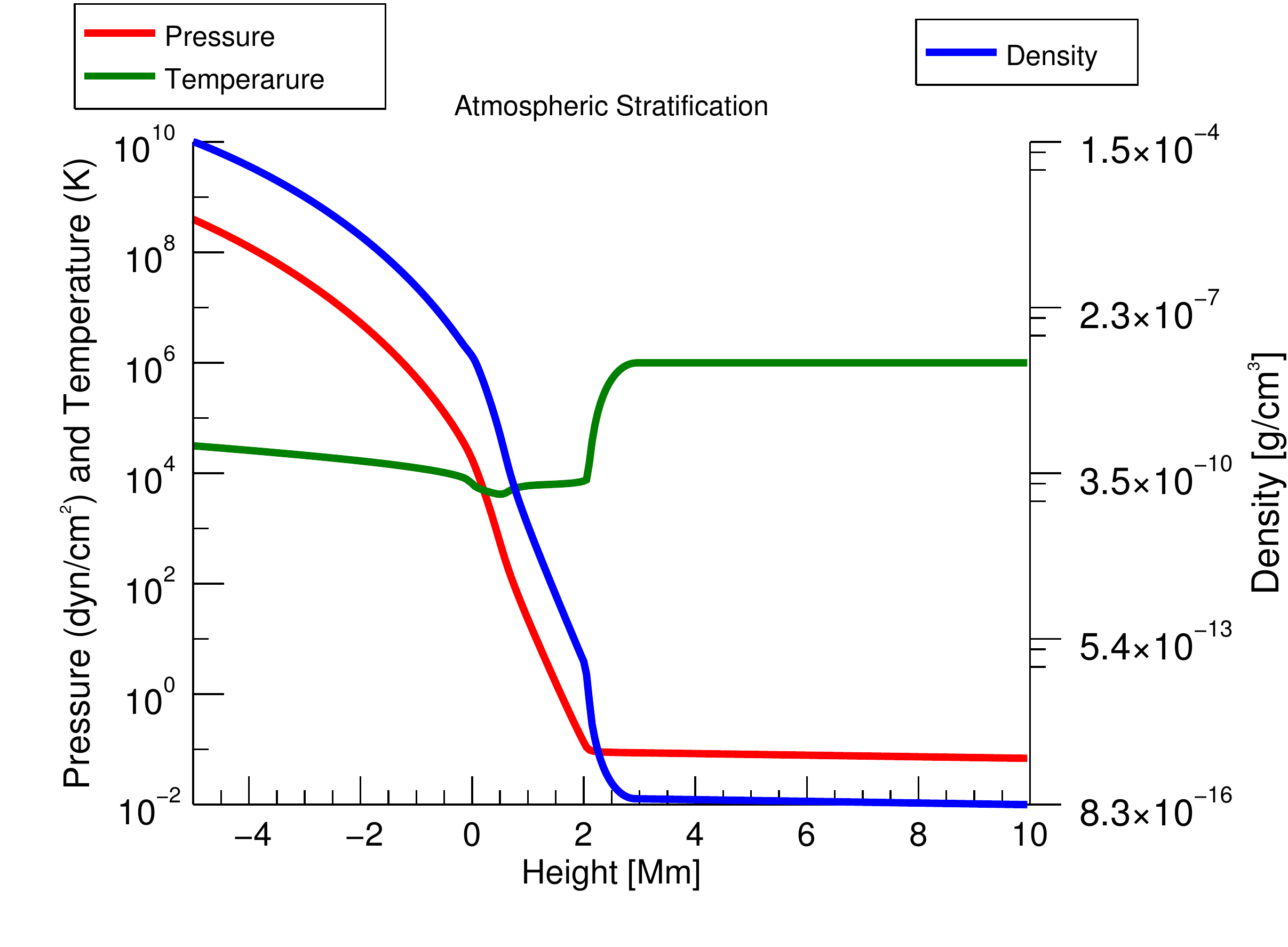}
\caption{Temperature (green line), density(blue line) and pressure (red line) distributions. The left vertical axis gives us the values for temperature and pressure, and the vertical axis on the right is the density scale.}
\label{fig:temperature}
\end{figure*}

To get the hydrostatic equilibrium we start by defining the temperature structure as a function of the vertical coordinate $z=\{-5,10\}$ Mm. For that, we join three models. The sub-photospheric layers are described by the convectively stabilized model by \citet{Parchevsky+Kosovichev2007}. The main reason to choose this model is because no convective instability is developed in response to the perturbation, allowing to study waves separate from convection \citep[see][]{Khomenko+etal2009}. The photosphere and chromosphere are described by VALC \citep{Vernazza+Avrett+Loeser1981} model for quiet Sun regions. Finally we choose an isothermal corona with a temperature of one million Kelvin. The complete temperature distribution is shown in Figure \ref{fig:temperature} (green line).  Once the temperature stratification is established we calculate the scale height $H$, and then integrate the equation for pressure and recover the density distributions, as follows:
\begin{eqnarray} \label{eq:HS}
&&H=\frac{R_{\rm gas}T_0}{g\mu} \\ %\nonumber
&&\frac{dp_{0}(z)}{dz}  + \frac{p_0}{H} =0 \\% \nonumber
&&\rho_0= \frac{p_0}{gH}
\end{eqnarray}
using the ideal gas equation.  With this, we ensure that the atmosphere is in hydrostatic equilibrium. The resulting distributions of pressure and density are given in Fig. \ref{fig:temperature}. A plane-parallel atmosphere is built by replicating the obtained vertical stratifications. In the above equations we denote the equilibrium variables by a subscript ``0''; R$_{\rm gas}$ is the gas constant and $\mu$ is the mean atomic weight.  We set $\mu$=1 at the lower layers of the atmosphere where we can consider that the gas is made of neutral hydrogen, and $\mu$=1/2 at higher layers where the plasma is fully ionized.

%%%%%%%%%%%%%%%%%%%%%%%%%%%%%%%%%%%%%%%%%%%%%%%
%%%%%%%%%%%%%%%%%%%%%%%%%%%%%%%%%%%%%%%%%%%%%%%

%%%%%%%%%%%%%%%%%%%%%%%%%%%%%%%%%%%%%%%%%%%%%%%
\subsubsection{Magnetostatic model}
%%%%%%%%%%%%%%%%%%%%%%%%%%%%%%%%%%%%%%%%%%%%%%%

\begin{figure} 
\centering
\includegraphics[width=9cm]{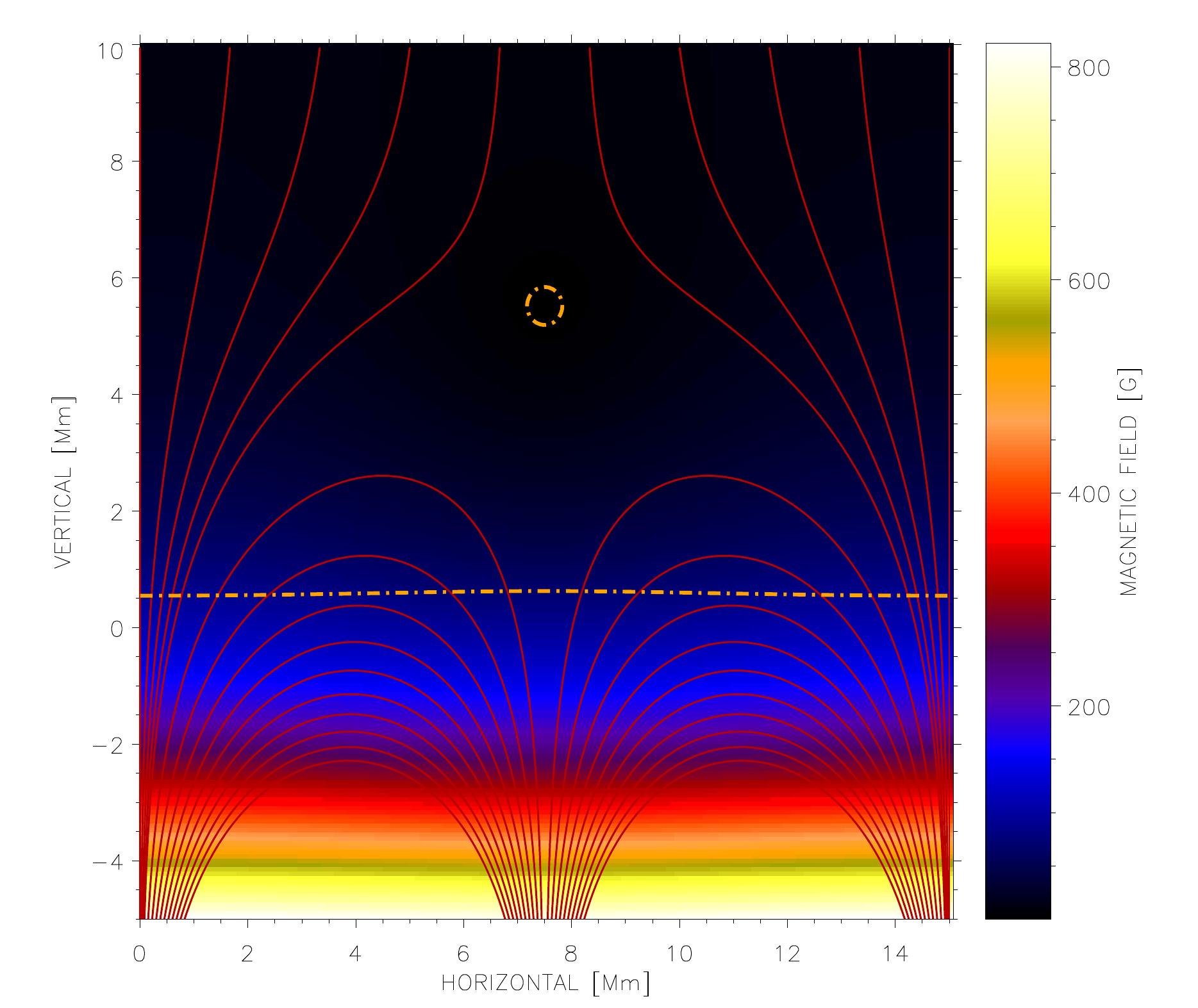}
\caption{Magnetic field strength distribution from below the photosphere ($-5$ Mm) to the corona (10 Mm). The vertical axis gives the height in Mm and the horizontal axis gives the horizontal size in Mm. Height z=0 represents the base of the photosphere. The red lines are the magnetic field lines and the yellow dashed lines are the $\beta$ =1 contours (one on the photosphere and the other one around the null point). } \label{fig:magfield}
\end{figure}

The magnetostatic model is given by an arcade shaped magnetic field plus a constant component:
\begin{eqnarray} \label{eq:mag_field}
B_{0x} &=& B_{00}\exp(-kz)\sin(kx) \\ %\nonumber
B_{0z} &= &B_{u}+B_{00}\exp(-kz)\cos(kx) 
\end{eqnarray}
where B$_{00}$ and B$_{u}$ are the magnetic field strengths in the photosphere and corona, respectively, $k$ defines the spatial (horizontal and vertical) scales and $x$ and $z$ are the horizontal and vertical directions, respectively. We choose B$_{00}$ = 100 G and B$_{u}$=10 G, so we are working with a weak magnetic field region. Equations \ref{eq:mag_field} defines a magnetic field configuration composed by two vertical flux tubes separated by an arcade-shaped magnetic field, as drawn in Figure \ref{fig:magfield}. Choosing higher values for the photospheric and coronal magnetic field has the disadvantage of significantly reducing the computational time steps due to the high values of the Alfv\'en speed in the computational domain. Higher values of $B_{00}$ and $B_u$ would lower the eventual position of the $\beta=1$ contour, but would not change the overall picture of the wave propagation and conversion. Table \ref{tab:velocities} gives sound and Alfv\'{e}n speeds for the different layers in the equilibrium atmosphere.

%%%%%%%%%%%%%%%%%%%%%%%%%%%%%%%%%%%%%%%%%%%%%%%
\begin{table*}
\begin{center}
\caption[]{\label{tab:velocities}
          { Alfv\'{e}n and sound speeds in the different layers of the equilibrium atmosphere.}}
\begin{tabular}{cccc}
\hline 
  		  & Photosphere  & Chromosphere & Corona \\
 \hline
 c$_{s0}$ & 8 km/s   	 &   10 km/s	    & 100 km/s	\\
 v$_{a0}$ & 2 km/s	 	 &   100 km/s	& 1000 km/s	  \\
\hline
\end{tabular} \label{tab:velocities}
\end{center}
\end{table*}
%%%%%%%%%%%%%%%%%%%%%%%%%%%%%%%%%%%%%%%%%%%%%%%

%%%%%%%%%%%%%%%%%%%%%%%%%%%%%%%%%%%%%%%%%%%%%%%
%%%%%%%%%%%%%%%%%%%%%%%%%%%%%%%%%%%%%%%%%%%%%%%

Our field configuration includes a null point in the corona. In this point the magnetic field strength is mathematically zero and the plasma $\beta$ goes to infinity. This null point changes drastically the behaviour of the waves as they propagate near it. As the magnetic field strength is zero, the Alfv\'{e}n speed is null, and therefore, only the acoustic waves can physically pass through it. 

%%%%%%%%%%%%%%%%%%%%%%%%%%%%%%%%%%%%%%%%%%%%%%%
\subsection{Simulations setup}
%%%%%%%%%%%%%%%%%%%%%%%%%%%%%%%%%%%%%%%%%%%%%%%

We analyse three simulation runs, performed with the same model for the equilibrium atmosphere, but with a different driving of the waves. The cases include: (1) vertical driving, when the analytical solution for an acoustic wave with a period of 200 sec (fast mode) is applied at the lower boundary of the simulation domain; (2) horizontal driving, when the analytical solution for the slow magnetic wave with a period of 300 sec is applied at the lower boundary; (3) instantaneous pressure pulse below the photosphere. Table \ref{tab:setup} gives a summary for these runs, including the section and the figures where the results are presented. We choose the periodicities of 200 and 300 sec to study the behaviour of waves with frequencies below and above the cut-off frequency, typical for the solar conditions. The cut-off frequency layer for 300 sec period waves is located below the photosphere, those for 200 sec waves is at the temperature minimum. 

%%%%%%%%%%%%%%%%%%%%%%%%%%%%%%%%%%%%%%%%%%%%%%%
\begin{table*}
\begin{center}
\caption[]{\label{tab:simulations}
          { Summary of the simulation runs}}
\begin{tabular}{cccccc}
\hline 
Section  & Driving                       & Amplitude & Period  & dx/dz/Duration & Figures \\
 \hline
 3.1        &  vertical, harmonic     & 10$^{-2}$\ms\  & 200 s & 75 km / 50 km / 3000 s  & 3--6 \\
 3.2        &  horizontal, harmonic &  0.2 \ms\          & 300 s & 75 km / 50 km / 3000 s  & 7--10 \\
 3.3        &  pressure pulse          & 10$^{-5}$p$_{0}$  & $-$     & 75 km / 50 km / 3000 s  & 11--13 \\ 
\hline
\end{tabular} \label{tab:setup}
\end{center}
\end{table*}
%%%%%%%%%%%%%%%%%%%%%%%%%%%%%%%%%%%%%%%%%%%%%%%

%%%%%%%%%%%%%%%%%%%%%%%%%%%%%%%%%%%%%%%%%%%%%%%
\subsection{Boundary conditions}
%%%%%%%%%%%%%%%%%%%%%%%%%%%%%%%%%%%%%%%%%%%%%%%

To avoid spurious reflections of waves at the upper boundaries of the domain, we add a Perfectly Matched Layers \citep[PML][]{Berenger1994} of 20 grid points, see \citet{Felipe+etal2010} for details. The PML is specially designed to absorb the perturbations  that reach the borders of the domain, and was shown to successfully perform in the simulations of MHD waves \citep[see][]{Parchevsky+Kosovichev2008, Khomenko+etal2008, Felipe+etal2010, Hanasoge+etal2010}. In the simulations with a periodic driver, the lower boundary is replaced by the analytical solution for a given type of wave. In the simulations with an instantaneous pulse, the lower boundary contains the PML layer as well. In the horizontal direction, boundary conditions for periodic variations are used. 

%%%%%%%%%%%%%%%%%%%%%%%%%%%%%%%%%%%%%%%%%%%%%%%
\section{Driving waves with linear perturbations}
%%%%%%%%%%%%%%%%%%%%%%%%%%%%%%%%%%%%%%%%%%%%%%%

This section presents the results of the different simulation runs. Despite the code can handle non-linearities, the amplitude of the perturbation is sufficiently small and the wave propagation is kept in the linear regime in all the cases, so no shocks are formed. This is done on purpose because our intention is to study the wave propagation, conversion, refraction and reflection in the linear regime. To better understand the results of the simulations, we briefly summarize the nomenclature, and some basics of the mode conversion theory.

%%%%%%%%%%%%%%%%%%%%%%%%%%%%%%%%%%%%%%%%%%%%%%%
\subsection{Wave transformation}
%%%%%%%%%%%%%%%%%%%%%%%%%%%%%%%%%%%%%%%%%%%%%%%

Dashed-dotted curves in Figure \ref{fig:magfield} show the $\beta =1$ contours in our equilibrium model, one in the photosphere and another one around the null point. The plasma $\beta$ gives us the information of whether the atmosphere is magnetically dominated or not. Strictly speaking, the plasma $\beta$ is defined as $\beta=P_{\rm gas}/ P_{\rm mag}$, but from the point of view of the wave propagation, it is more convenient to define it as a squared ratio of the characteristic speeds, i.e. $\beta=c_s^{2}/ v_a^2$. The difference between both definitions is a factor $\gamma/2$, which in our case is given by 5/6. This value is close to unity and, therefore, almost imperceptible in the figure or in wave effects. We will use the latter definition through the paper. In the layer with $\beta \approx 1$ (equipartition layer), transformation of different wave modes can occur. Since our simulations are 2D, the only possible wave modes are fast and slow magneto-acoustic-gravity modes, so we exclude the conversion to Alfv\'{e}n mode from our current analysis.

The propagation of the fast and slow magneto-acoustic-gravity modes is in general rather complex, except in cases with very simple magnetic topologies. Under a mathematical point of view, the analysis of the behaviour of waves comes in a natural way if done according to the fast or slow nature of their propagation speed, regardless of their acoustic or magnetic properties. However, for the interpretation of observations, it is in general more convenient to take into account the physical nature of the main restoring force. It is unfortunate that the terms wave {\it transmission} and {\it conversion} have different meanings in these two descriptions. Since this paper focusses on the interpretation of numerical simulations, we have preferred to follow the physical convention. This way, we will consider that a wave is transmitted when the nature of its main restoring force does not change when crossing a given layer, keeping its mainly acoustic/magnetic nature (independently of whether in the mathematical description it may change from fast to slow or viceversa). Similarly, we will use the term wave conversion when the nature of the main restoring force changes from acoustic/magnetic to magnetic/acoustic (keeping the fast or slow character or the wave). 

Depending on the magnetic field inclination, the conversion and transmission are usually only partial \citep{Zhugzhda+Dzhalilov1982, Cally2005, Cally2006, Schunker+Cally2006, Khomenko+Collados2006, Khomenko+Cally2012}, hence part of the wave keeps its acoustic/magnetic nature and part of it changes. The transmission efficiency depends on the attack angle, i.e., the angle between the magnetic field lines and the wave vector in the layer of $\beta =1$ \citep{Cally2006}: 

\begin{equation} \label{eq:convers}
T = \exp \Big [- \frac{K \pi \sin^{2}\alpha}{\mid (d/ds) (c_{s}^{2}/v_{a}^{2}) \mid}\Big]_{c_{s}=v_{a}}  \,,
\end{equation} 
where $\alpha$ is the attack angle and $s$ is the distance along the direction of $\vec{k}$ and $K=|\vec{k}|$. $c_{s}/v_{a}$ is the ratio of the speed of sound and the Alfv\'{e}n speed. The transmission is complete when the magnetic field and the direction of the wave propagation are aligned, $\alpha=0$. 
In this case, the conversion coefficient, $1-T$, is zero. When $\alpha$ $\neq$ 0, then the transmission decreases with increasing k, that is, with increasing wave frequency.

%%%%%%%%%%%%%%%%%%%%%%%%%%%%%%%%%%%%%%%%%%%%%%%
\subsection{Vertical periodic driver of 200 seconds}
%%%%%%%%%%%%%%%%%%%%%%%%%%%%%%%%%%%%%%%%%%%%%%%

In this run we drive waves by applying a vertical periodic perturbation in velocity and thermodynamic variables at the bottom of the domain and let the perturbation propagate a total duration of 3000 seconds. This duration is long enough for the waves to fill the full domain and reach a stationary motion. The direction of propagation of the perturbation in this case is nearly parallel to the magnetic field lines, so acoustic like fast mode waves are mostly generated. We use an analytical solution for an acoustic-gravity wave in an isothermal atmosphere, see \citet{Mihalas+Mihalas1984}. The perturbations are given by:
\begin{eqnarray} \label{eq:initial}
\delta v_{z} &  =&  V_0 \exp\left( \frac{z}{2H}+k_{zi}z \right)\sin(\omega t - k_{zr} z) \\ %\nonumber
\frac{\delta p}{p_0} & =&  V_0|P| \exp\left( \frac{z}{2H}+k_{zi}z \right)\sin(\omega t - k_{zr} z + \phi_P) \\ %\nonumber
\frac{\delta\rho}{\rho_0} & =& V_0 |R| \exp\left( \frac{z}{2H}+k_{zi}z \right)\sin(\omega t - k_{zr} z + \phi_R)
\end{eqnarray}
where $V_0=10^{-2}$ \ms\ is the amplitude of the velocity, see Table \ref{tab:setup}, and $H$ is the pressure scale height. Perturbations in the horizontal velocity and the magnetic field are null. k$_{zr}$ and k$_{zi}$ are the real and complex parts of the vertical wave number,respectively. Given the wave frequency $\omega$, this is found from the dispersion relation:
\begin{equation} \label{eq:kz}
k_{z} = k_{zr} + i k_{zi} =  \frac{\sqrt{\omega^{2} - \omega_{c}^{2}}}{c_{s}} 
\end{equation}
where $\omega_{c} = \gamma g / 2 c_{s} $ is the isothermal acoustic cut-off frequency. 
 
The relative amplitudes and phase shifts between the perturbations are given by
\begin{eqnarray}
|P| &=& \frac{\gamma}{\omega}\sqrt{k_{zr}^2 + \left( k_{zi} + \frac{1}{2H}\frac{(\gamma - 2)}{\gamma} \right)^2}\\
|R| &=& \frac{1}{\omega}\sqrt{k_{zr}^2 + \left( k_{zi} - \frac{1}{2H} \right)^2}
\end{eqnarray}
\begin{eqnarray}
\phi_P &=& \arctan\left( \frac{k_{zi}}{k_{zr}} + \frac{1}{2Hk_{zr}}\frac{(\gamma - 2)}{\gamma}   \right)\\
\phi_R &=&\arctan\left(  \frac{k_{zi}}{k_{zr}}  - \frac{1}{2Hk_{zr}} \right)
\end{eqnarray}

The perturbation is imposed in the interval $z=\{-5,-4.65\}$ Mm, i.e. occupies 7 grid points. 

As is frequently done in MHD wave studies, to distinguish between the modes in the simulations, we project the velocities into the directions parallel and perpendicular to the local magnetic field, defined as:
\begin{eqnarray} \label{eq:projections}
e_{\rm long}&=& \sin \theta e_x + \cos \theta e_z \\
e_{\rm trans}&=& -\cos \theta e_x + \sin \theta e_z
\end{eqnarray}
where $\theta$ is the inclination of the magnetic field. This way, in the $\beta \ll 1$ atmosphere, the acoustic slow mode is visible in the longitudinal velocity (propagation parallel to the magnetic field); and the fast magnetic mode is essentially transverse to the magnetic field. Note that the projections do not help to distinguish between the modes in the $\beta \gg 1$ region. 

%%%%%%%%%%%%%%%%%%%%%%%%%%%%%%%%%%%%%%%%%%%%%%%
\begin{figure}
\centering
\includegraphics[width=6.9cm]{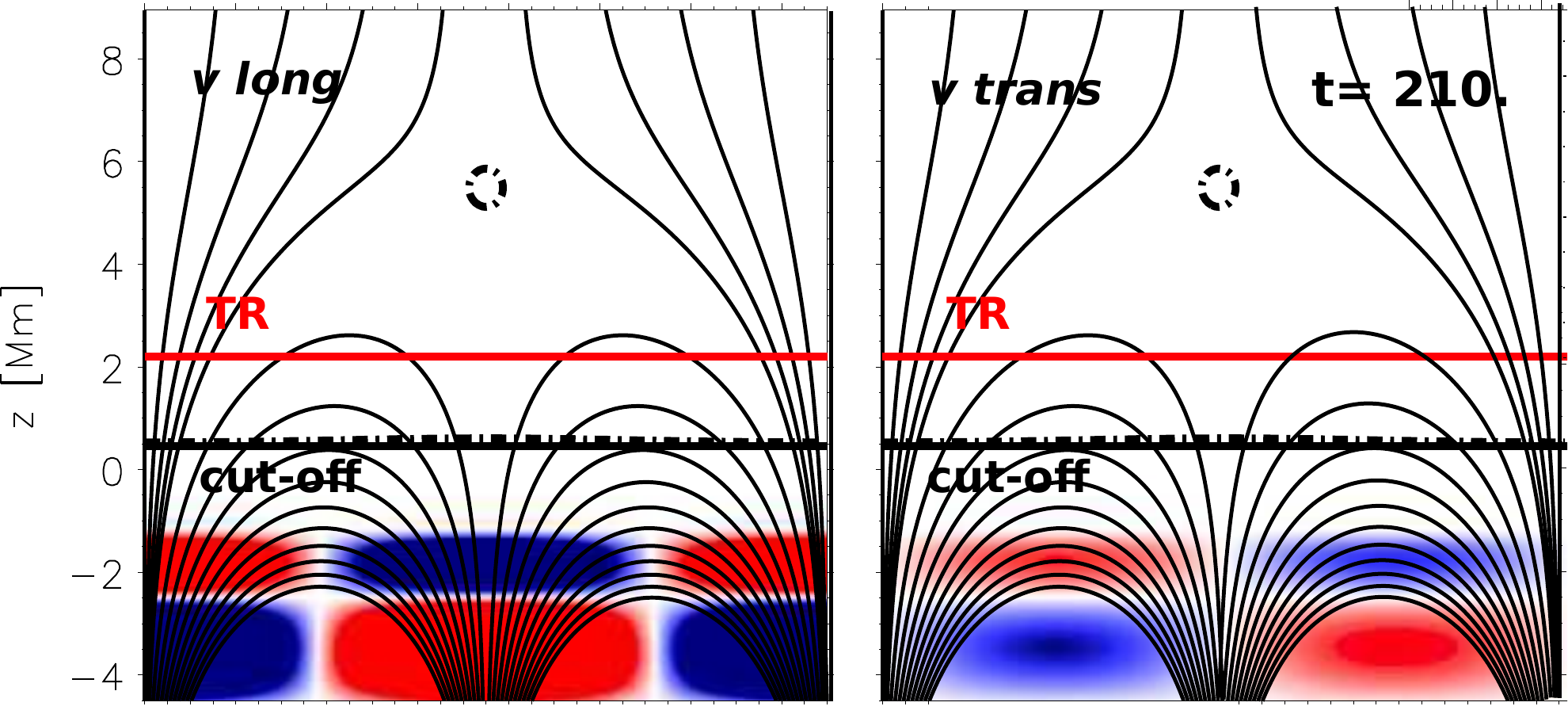}
\includegraphics[width=6.9cm]{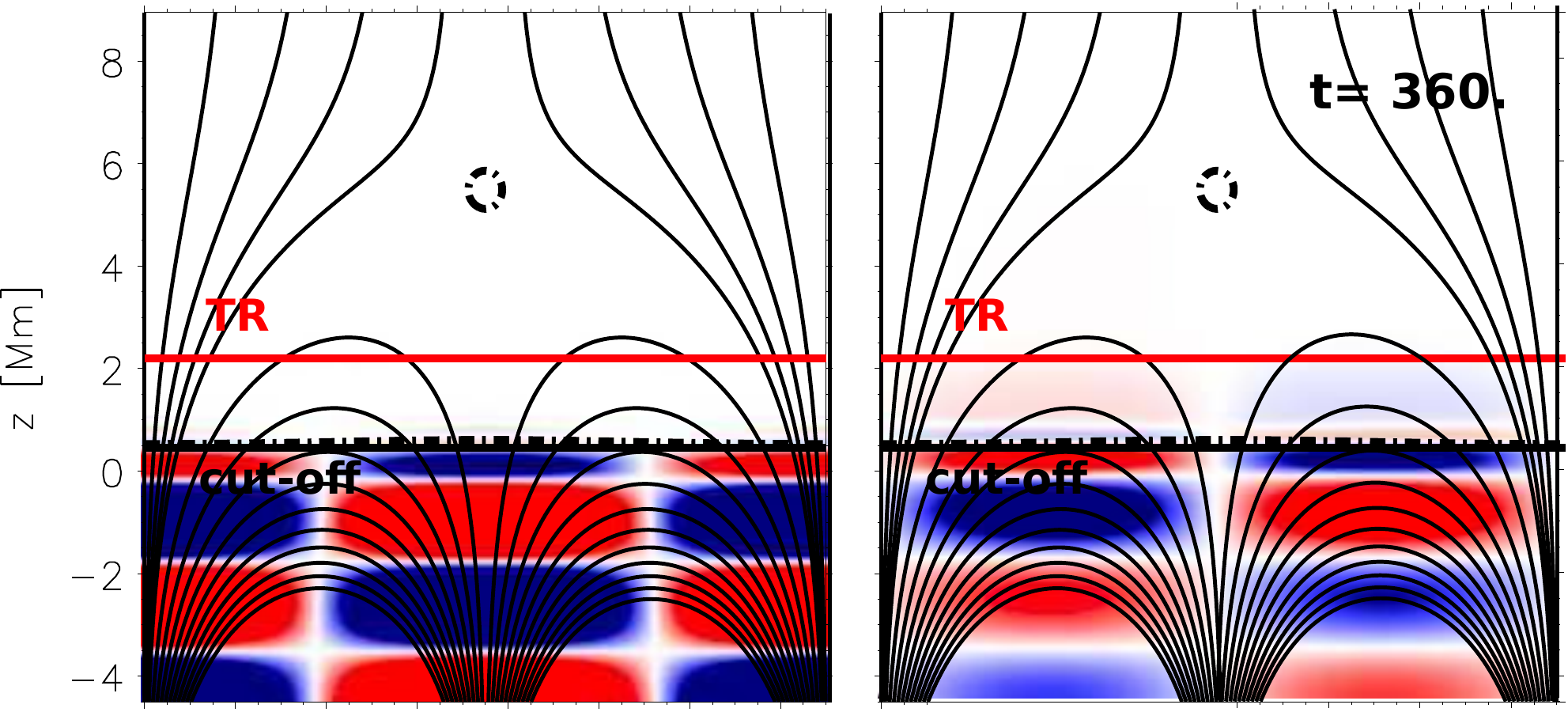}
\includegraphics[width=6.9cm]{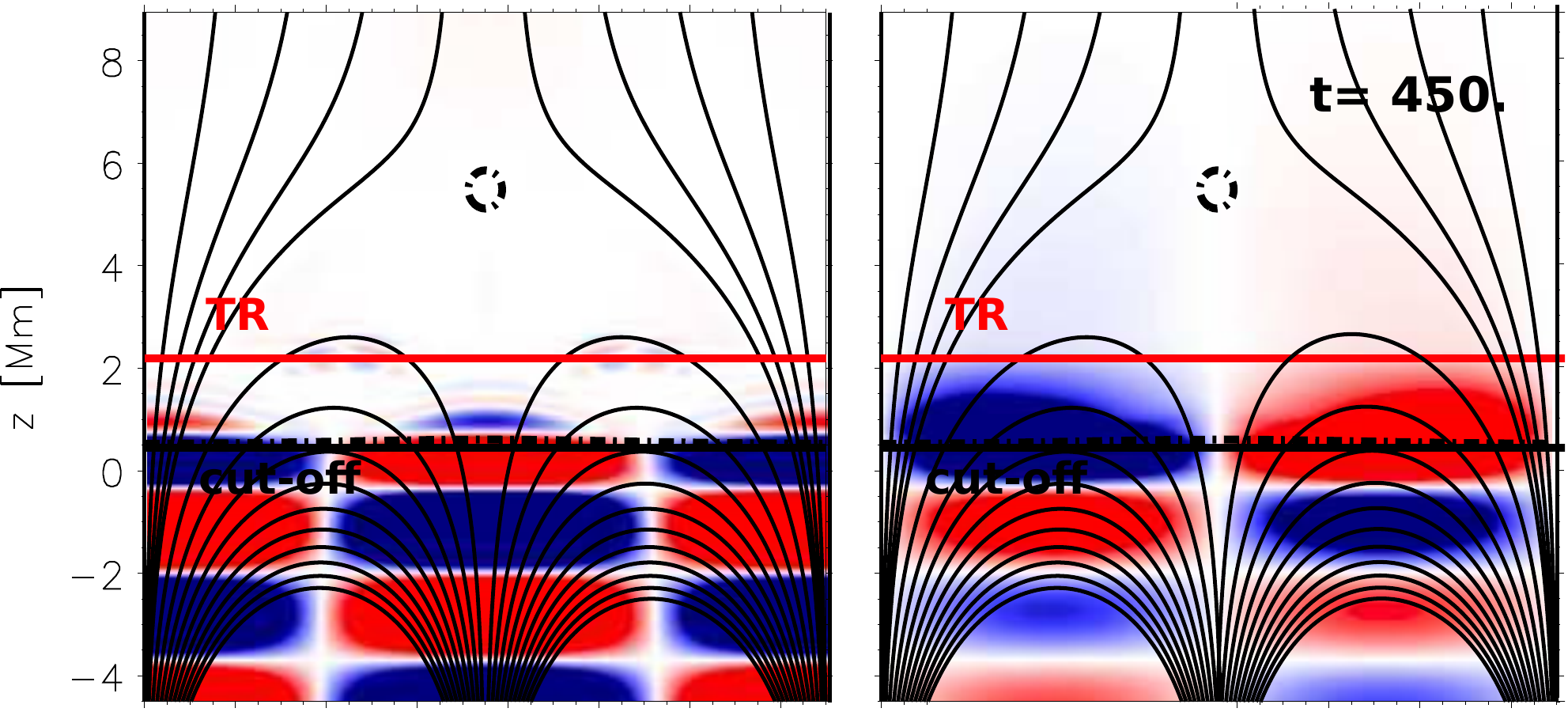}
\includegraphics[width=6.9cm]{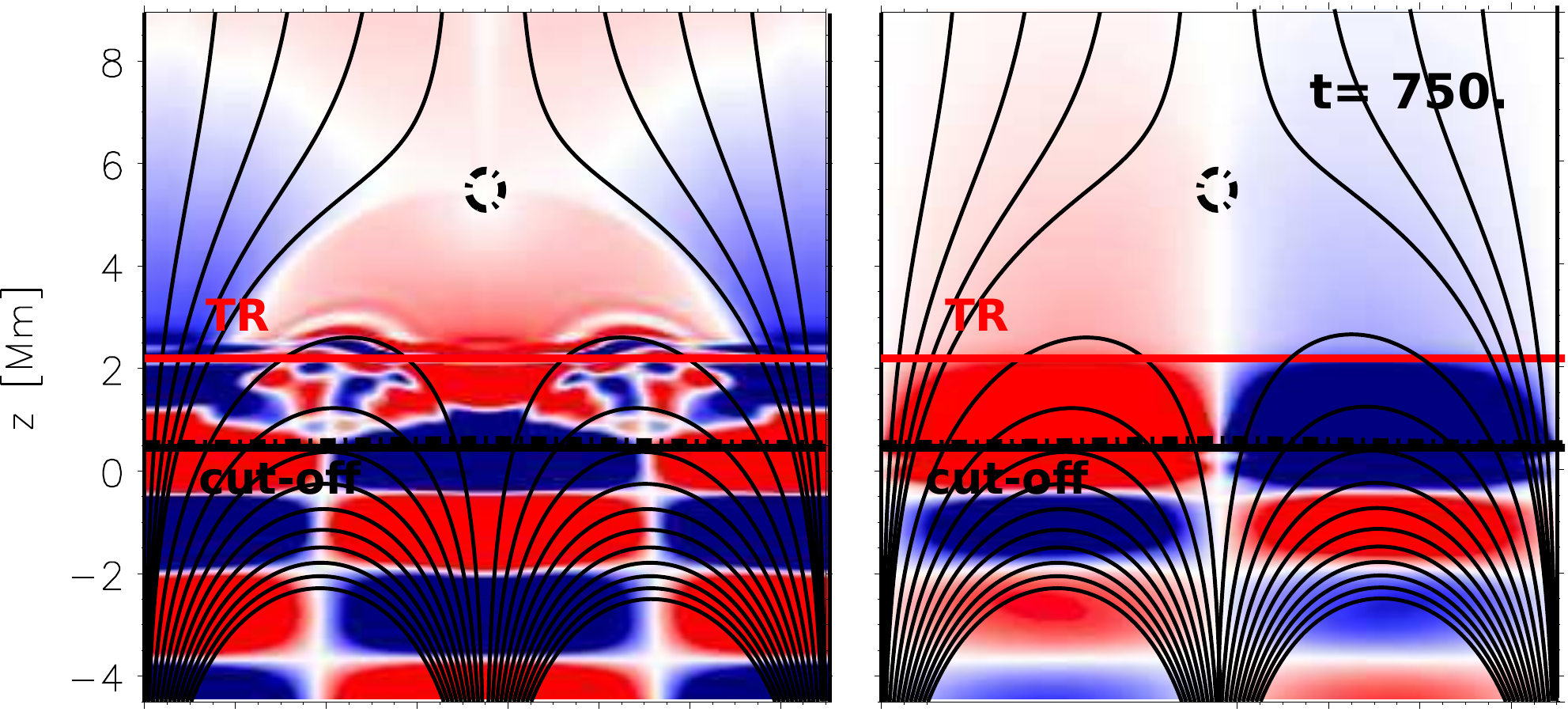}
\includegraphics[width=6.9cm]{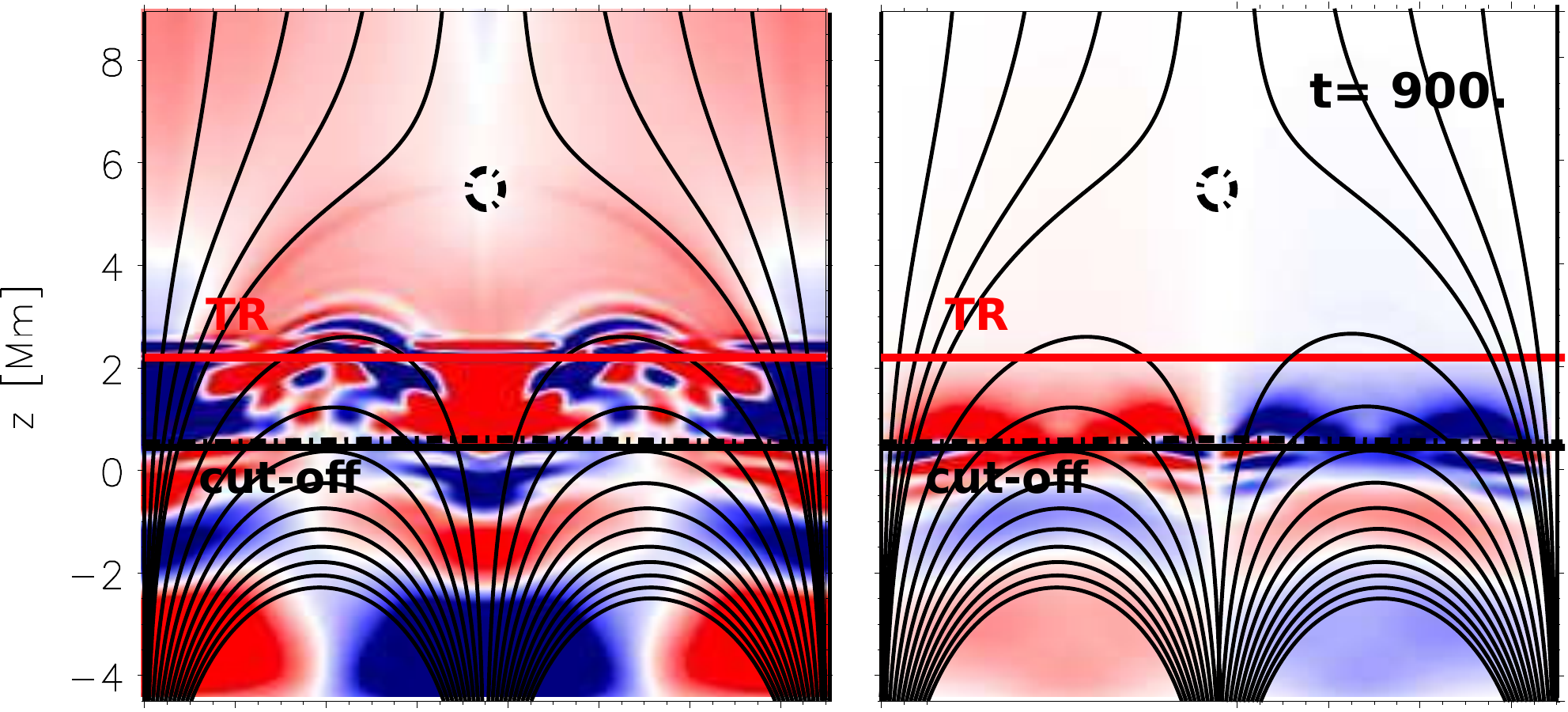}
\includegraphics[width=6.9cm]{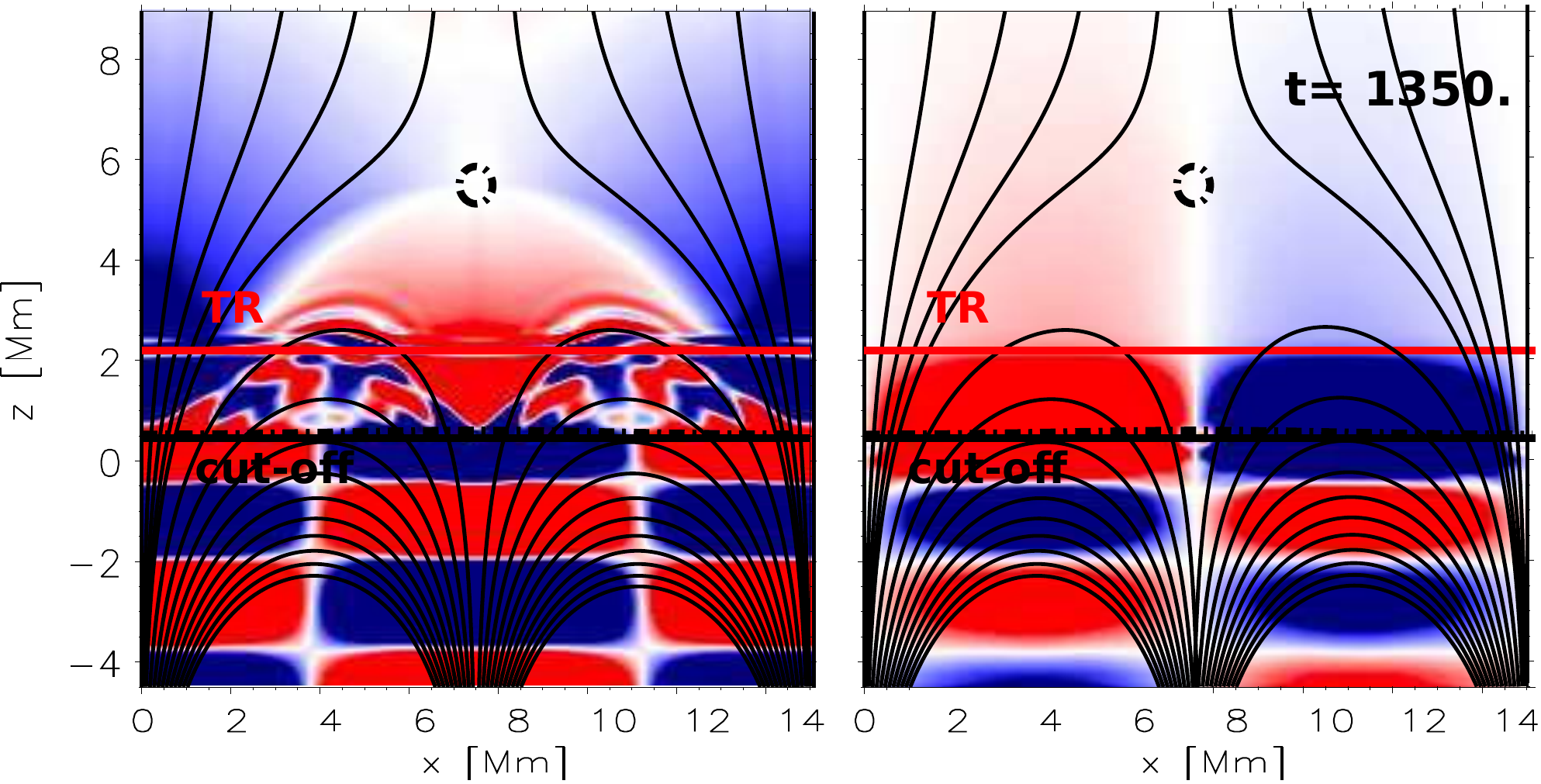}
\includegraphics[width=7.0cm]{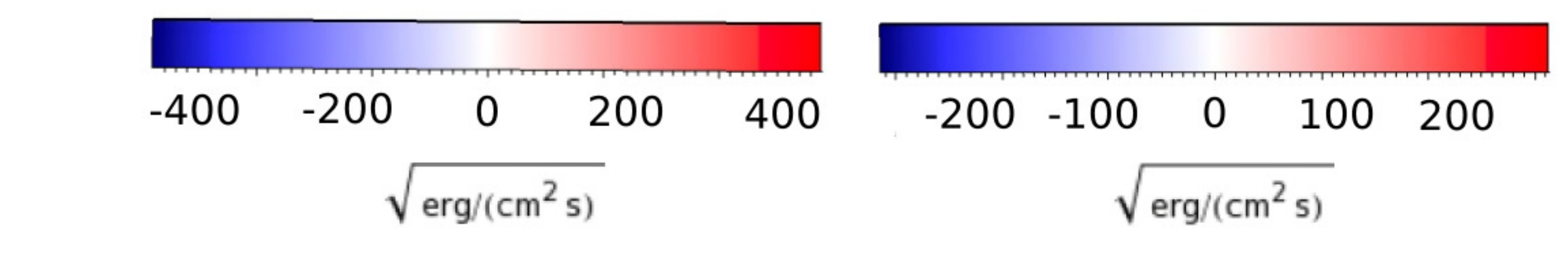} 
\caption{Longitudinal (left) and transverse (right) velocities multiplied by $\sqrt{\rho_{0}c_{s0}}$ and $\sqrt{\rho_{0}v_{A0}}$, respectively, for the simulation run with a vertical periodic driver. The black solid line is the location of the cut-off frequency for the 200 second period waves, the dashed black lines are the $\beta $=1 contours and the red line shows the transition region. The magnetic field lines are inclined black lines. Positive velocity (red color) is an upflow. Note that the horizontal variations in the longitudinal and transverse velocities in the bottom layers where the driver is imposed, despite the velocity perturbation is constant in the horizontal direction (see Eq. \ref{eq:initial}). These changes are caused by the projection of the velocity on the directions parallel and perpendicular to the magnetic field, which has a varying inclination. The movie of the wave propagation is available in the online version of the paper.} \label{fig:vlvt}
\end{figure}
%%%%%%%%%%%%%%%%%%%%%%%%%%%%%%%%%%%%%%%%%%%%%%%

%%%%%%%%%%%%%%%%%%%%%%%%%%%%%%%%%%%%%%%%%%%%%%%
\subsubsection{Wave propagation}
%%%%%%%%%%%%%%%%%%%%%%%%%%%%%%%%%%%%%%%%%%%%%%%

Figure \ref{fig:vlvt} shows the longitudinal (left panels) and transverse (right panels) velocity projections, scaled with the factors $\sqrt{\rho_{0}c_{s}}$ and $\sqrt{\rho_{0}v_{a}}$, respectively. Such scaling gives an approximation for the energies contained in the different wave components. Some important layers are also indicated in each panel of this figure: $\beta$=1 contours (dashed black lines), the transition region (red line) and the location of the cut-off layer with a period of 200 seconds (black solid line), the same period as that of the driver. 

The magneto-acoustic fast and slow waves start propagating upwards through the layers below the photosphere, where plasma $\beta \gg 1$  (see the rows $t=210$ s and $t=360$ s). As the driver is parallel to the magnetic field at the foot points, the amplitude of the longitudinal velocity is larger than the transverse one. 

As the waves reach the equipartition layer at $t=450$ s, fast acoustic-like waves are partially converted into fast magnetic-like waves and partially transmitted into longitudinal slow acoustic waves.  These can be seen comparing the second (360 s) and third (450 s) rows of Fig. \ref{fig:vlvt}: in the second row the transverse component crosses the equipartition layer with a very small amplitude, but once the longitudinal component reaches the layer where $\beta = 1$ at 450 s, the amplitude of the transverse component grows considerably while the amplitude of the longitudinal waves decreases.

These waves continue propagating upwards until they reach the transition region (before 600 s). At this point, part of the wave energy is reflected and part is transmitted into the corona. The transverse velocity component suffers significant reflection from the transition region (right panels from 750 s to the end). The energy of the transverse component is mostly concentrated between the equipartition layer and the transition region. 

Some of the energy of the longitudinal waves is transmitted to the corona through the transition region. The behaviour of these waves, once they pass the transition region, depends on whether they propagate near the null point or far from it (see left panels of Fig. \ref{fig:vlvt} from t=750 s to the end). The waves propagating near the lateral borders of the domain, inside the almost vertical flux tubes, are longitudinal and continue propagating upwards along the field lines up to the upper corona. Waves propagating outside the vertical flux tubes reach the null point and this changes drastically their behaviour. These waves suffer partial conversion, from magnetic to acoustic and vice versa, and transmission, from fast to slow and vice versa, at the equipartition layer around the null point. The Alfv\'en speed is zero at the null, and therefore, only acoustic fast waves (inside the circle marked by the coronal $\beta=1$ contour) are able to cross it. The movie of the simulations available in the online version of the paper) makes apparent how the null point ``absorbs'' the waves and then ``ejects'' them away in all directions. This effect is mainly caused by the refraction of the fast magnetic waves due to the large gradient of the Alfv\'{e}n speed. Part of these fast waves propagate downward again through the transition region and equipartition layer to the lower atmosphere. The null point acts as a re-feeding of the atmosphere. 

Few minutes after the waves reach the upper boundary, the simulation enters into the stationary stage. This property will be used for the calculation of the mean wave energy fluxes, done in the section below.

%%%%%%%%%%%%%%%%%%%%%%%%%%%%%%%%%%%%%%%%%%%%%%%
\begin{figure}
\centering
\includegraphics[width=9cm]{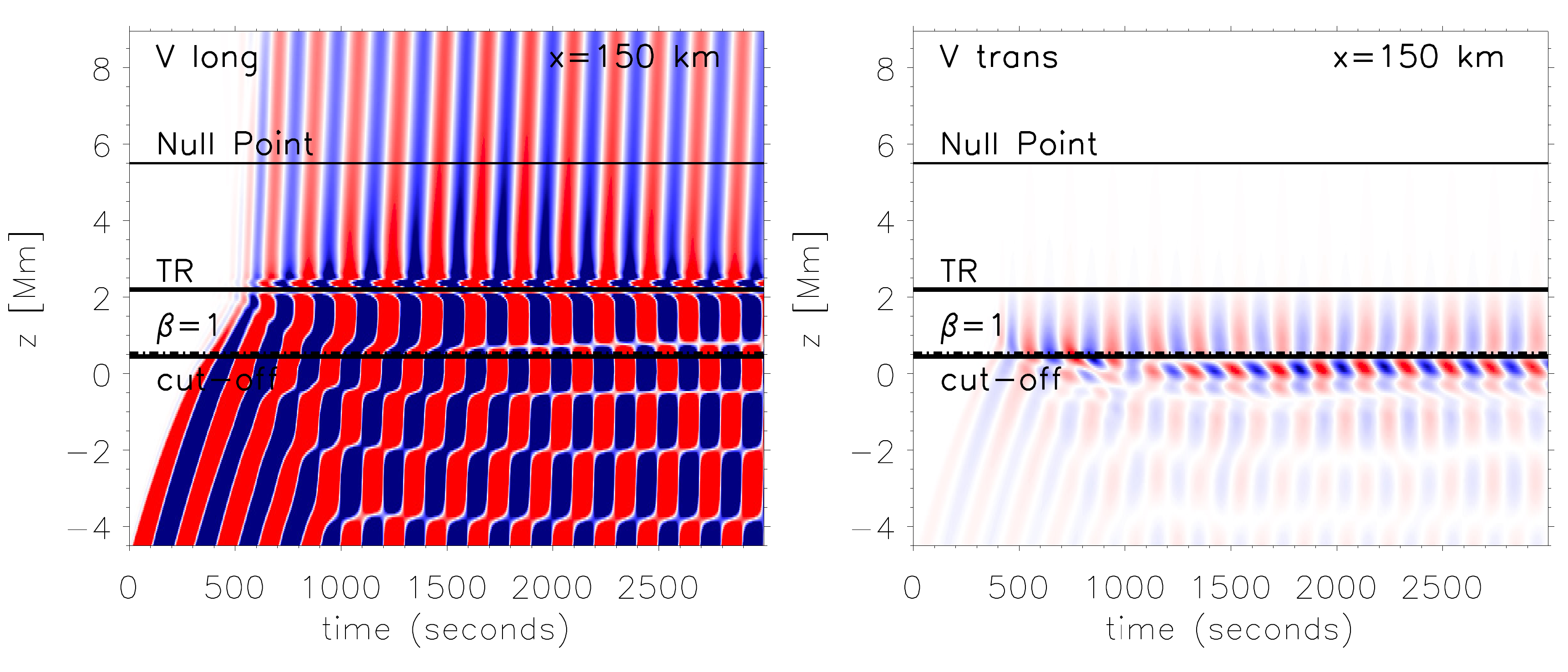}
\includegraphics[width=9cm]{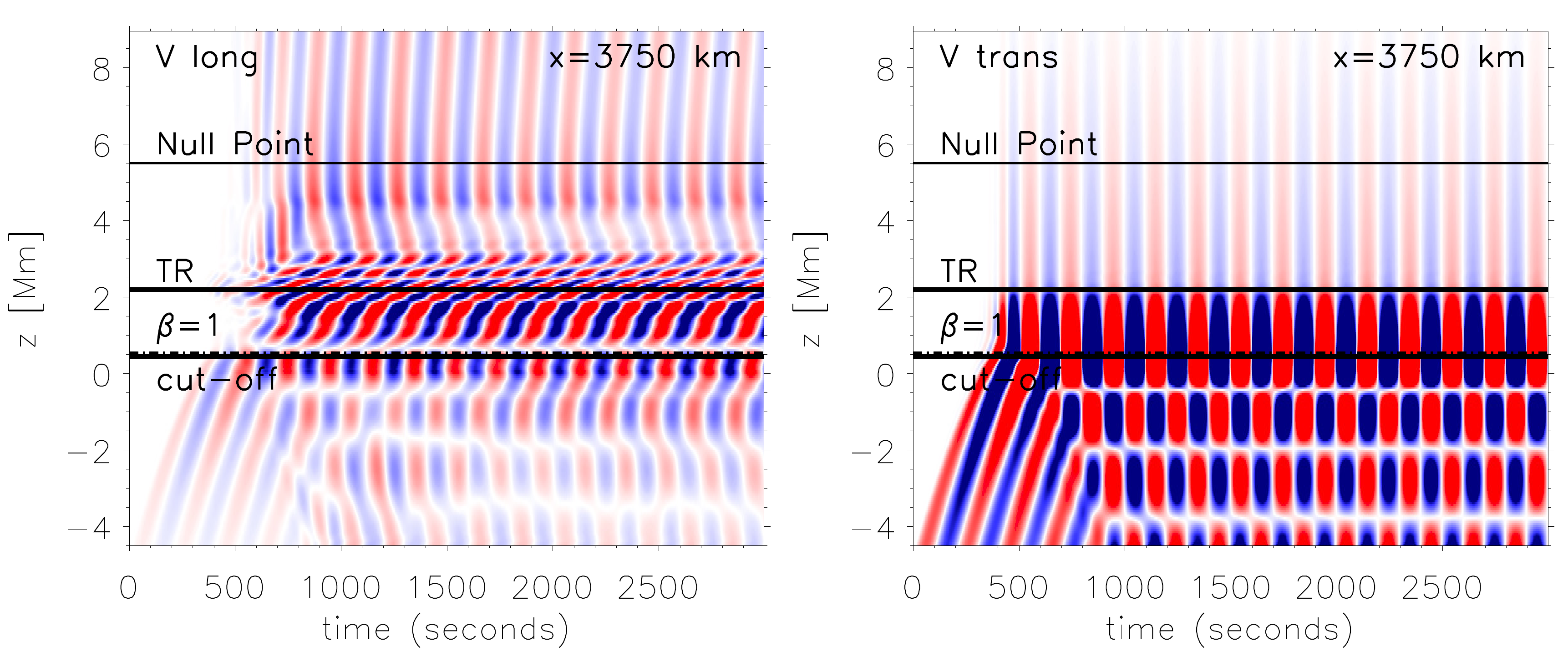}
\includegraphics[width=9cm]{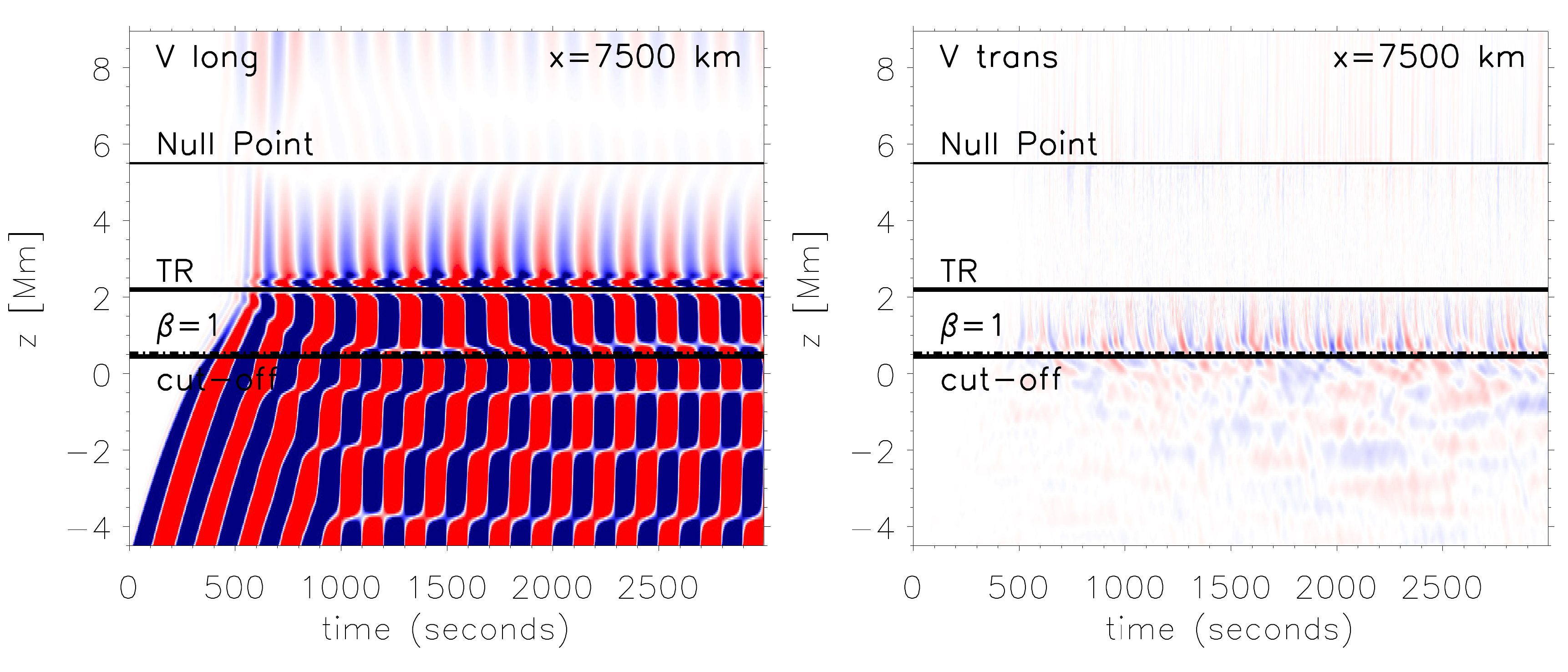}
\caption{Time-height diagram of $\sqrt{\rho_0 c_{s}}v_{\rm long}$ (left) and $\sqrt{\rho_0 v_{a}}v_{\rm trans}$ (right) for three selected horizontal positions in the vertical periodic driver simulations. Upper row: $x=150$ km, where the magnetic field is nearly vertical. Middle row: $x= 3750$ km, close to the center of the left arcade with nearly horizontal magnetic field. Bottom row:  $x=7500$ km close to the null point location. The color scale is the same as in Figure \ref{fig:vlvt}. We notice how the simulation reaches a stationary stage after about 1000 s.  } \label{fig:periodic200td}
\end{figure}
%%%%%%%%%%%%%%%%%%%%%%%%%%%%%%%%%%%%%%%%%%%%%%%

Time-height diagrams allow to appreciate more clearly the behaviour of waves described above, see Figure \ref{fig:periodic200td}.
There we plot the quantities $\sqrt{\rho_0 c_{s}}v_{\rm long}$ (left) and $\sqrt{\rho_0 v_{a}}v_{\rm trans}$ (right) at three different horizontal locations in the domain. The cuts are located at $x=150$ km inside the flux tube, where the magnetic field is nearly vertical, at $x= 3750$ km in the center of the left arcade and where the magnetic field is nearly horizontal, and at $x=7500$ km, in the middle of the domain where the null point is located. The inclination of ridges on this diagram reflects the wave propagation speed (more vertical means faster). Fig. \ref{fig:periodic200td} shows how the picture of wave interference changes at different magnetic field inclinations. 

The left panels demonstrate that waves propagating parallel to the nearly vertical magnetic field  at $x=150$ km and at $x=7500$ km travel almost undisturbed up to the transition region, where reflection is produced and an interference pattern is formed between the upward and downward propagating waves. As the attack angle is small at the $\beta = 1$ contour at these locations, the transmission from fast acoustic to slow longitudinal acoustic waves is almost total. Once the waves at $x=150$ km reach the transition region at about 600 s of the simulation time, they continue propagating with significant amplitude upwards (top left panel of Fig. \ref{fig:periodic200td}). Unlike that, waves propagating at $x=7500$ km are affected by the null point. One may appreciate at the bottom left panel, how the waves propagating downward, due to the refraction suffered near the null point, meet the waves going upwards and their amplitudes cancel out.

The behaviour of waves propagating at $x=3750$ km (middle panels of Fig. \ref{fig:periodic200td}), where the magnetic field is nearly horizontal below the transition region, is different from above. One can observe several small-scale wave components, with their energy located essentially between the $\beta=1$ contour and the transition region (middle left panel). These small-scale waves are a result of the mode conversion into slow modes propagating along the inclined field (their propagation speed is apparently small because of the longer distance they have to cross in the inclined fields). Fast magnetic modes have much larger propagation speeds and can be seen at the middle right panel as nearly vertical lines with a maximum power concentrated again in the chromosphere below the transition region. The reflection of longitudinal slow waves from the null point is also apparent at the middle left panel from the opposite inclination of the ridges above the transition region and below the null point. 

%%%%%%%%%%%%%%%%%%%%%%%%%%%%%%%%%%%%%%%%%%%%%%%
\subsubsection{Energy fluxes}
%%%%%%%%%%%%%%%%%%%%%%%%%%%%%%%%%%%%%%%%%%%%%%%

We calculate the mean acoustic and magnetic energy fluxes  as follows (Bray \& Loughhead, 1974):
\begin{eqnarray} \label{eq:fluxes}
{\vec{F}_{\rm ac}} & =  &\langle \delta p\,{\delta\vec{v}} \rangle \,,\\ 
{\vec{F}_{\rm mag}} & = &  \langle \delta\vec{B} \times ( \delta\vec{v} \times \vec{B}_0 ) \rangle /\mu_0\,.
\end{eqnarray}
In order to obtain the meaningful averages of the flux one needs to include at least several periods of the stationary regime of the simulations. For the simulations presented above, we considered that the stationary regime is maintained from 1000 to 3000 s of the simulation time. 

%%%%%%%%%%%%%%%%%%%%%%%%%%%%%%%%%%%%%%%%%%%%%%%
\begin{figure}
\centering
\includegraphics[width=9.0cm]{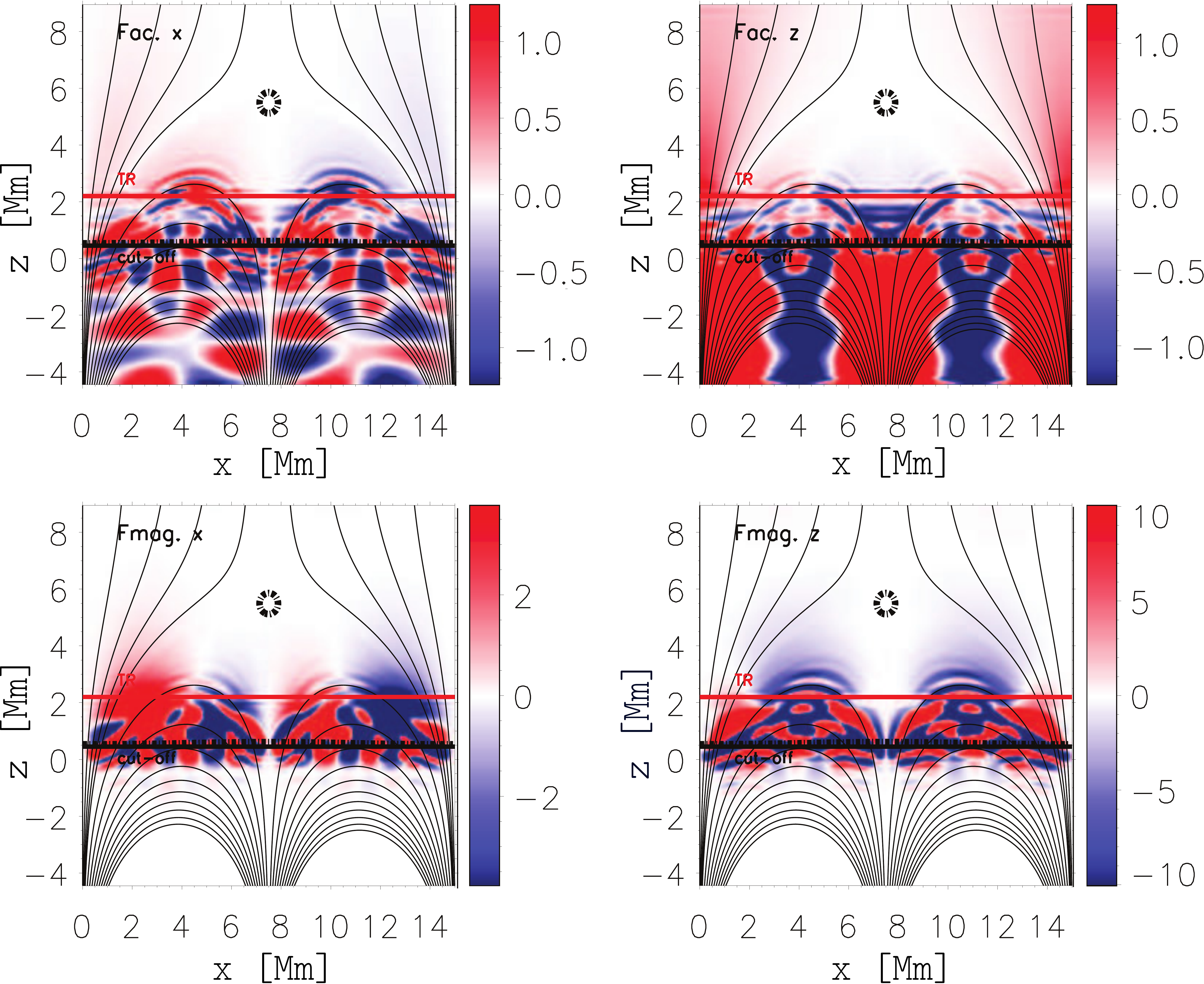}

\caption{Mean energy fluxes for the simulation with the vertical periodic driver. Top:  horizontal and vertical mean acoustic energy fluxes. Bottom:  horizontal and vertical magnetic mean energy fluxes. The lower black line is the acoustic cut-off frequency layer for 200 second period, the dashed lines are the $\beta =1$ contour, the red line is the transition region. The magnetic field lines are inclined black lines.  The units of the bar are 10$^6$ [erg/s/cm$^2$].}  \label{fig:periodic200flux}
\end{figure}
%%%%%%%%%%%%%%%%%%%%%%%%%%%%%%%%%%%%%%%%%%%%%%%

Figure \ref{fig:periodic200flux} shows the spatial distribution in the domain of the magnetic and acoustic, vertical and horizontal mean energy fluxes. It can be seen that almost all the magnetic energy (lower panels) is concentrated between the photosphere ($z=0$ km) and transition region ($z= 2100$ km), while the upward acoustic energy flux is significant in the corona inside the vertical flux tubes. The only significant upward energy flux in the corona is due to upward propagating longitudinal slow acoustic waves. No wave energy flux is present around the null point. This is happening because waves propagate in all directions around the null point and cancel out when waves with opposite directions meet.

\begin{figure}
\centering
\includegraphics[width=10cm]{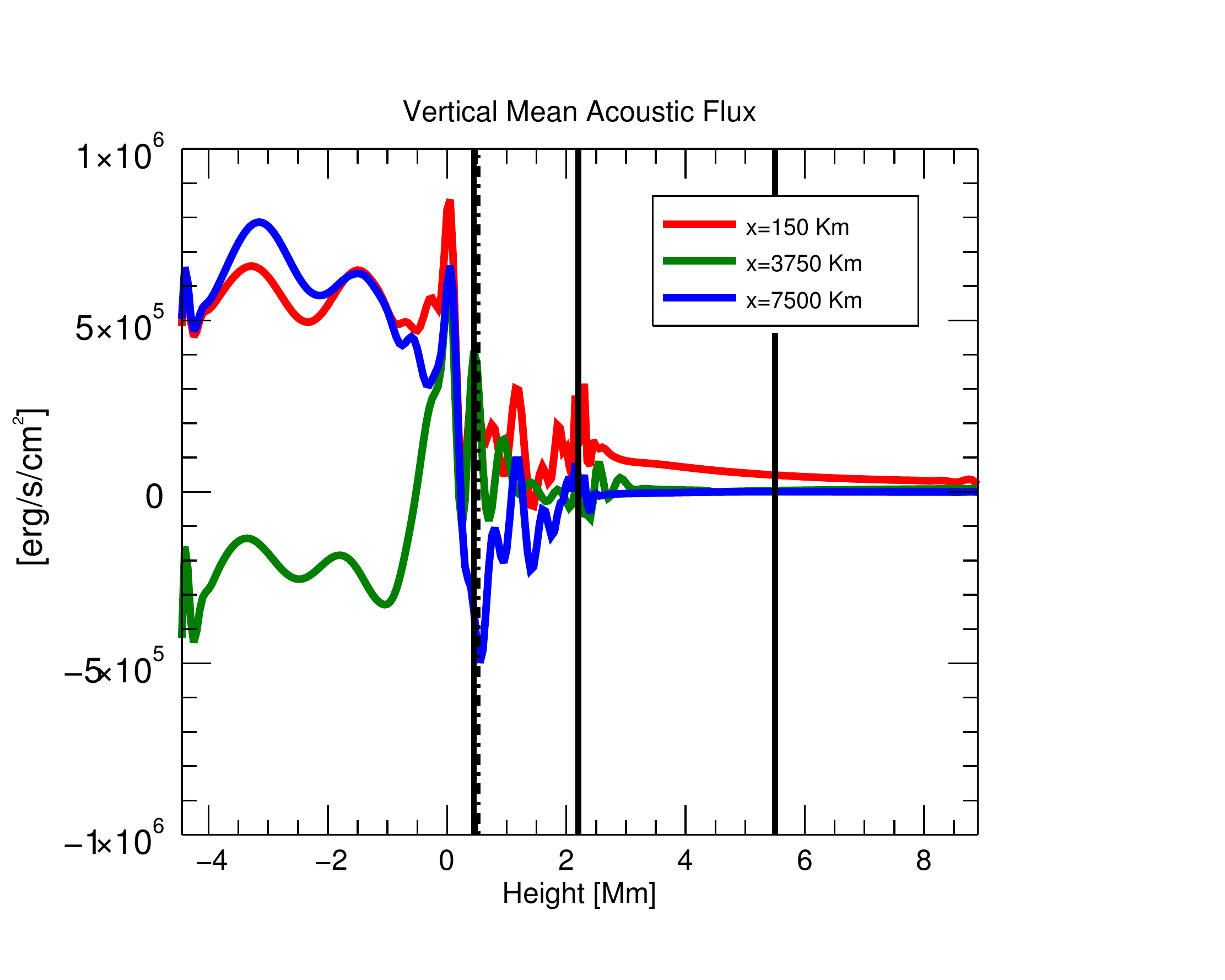} 
\includegraphics[width=10cm]{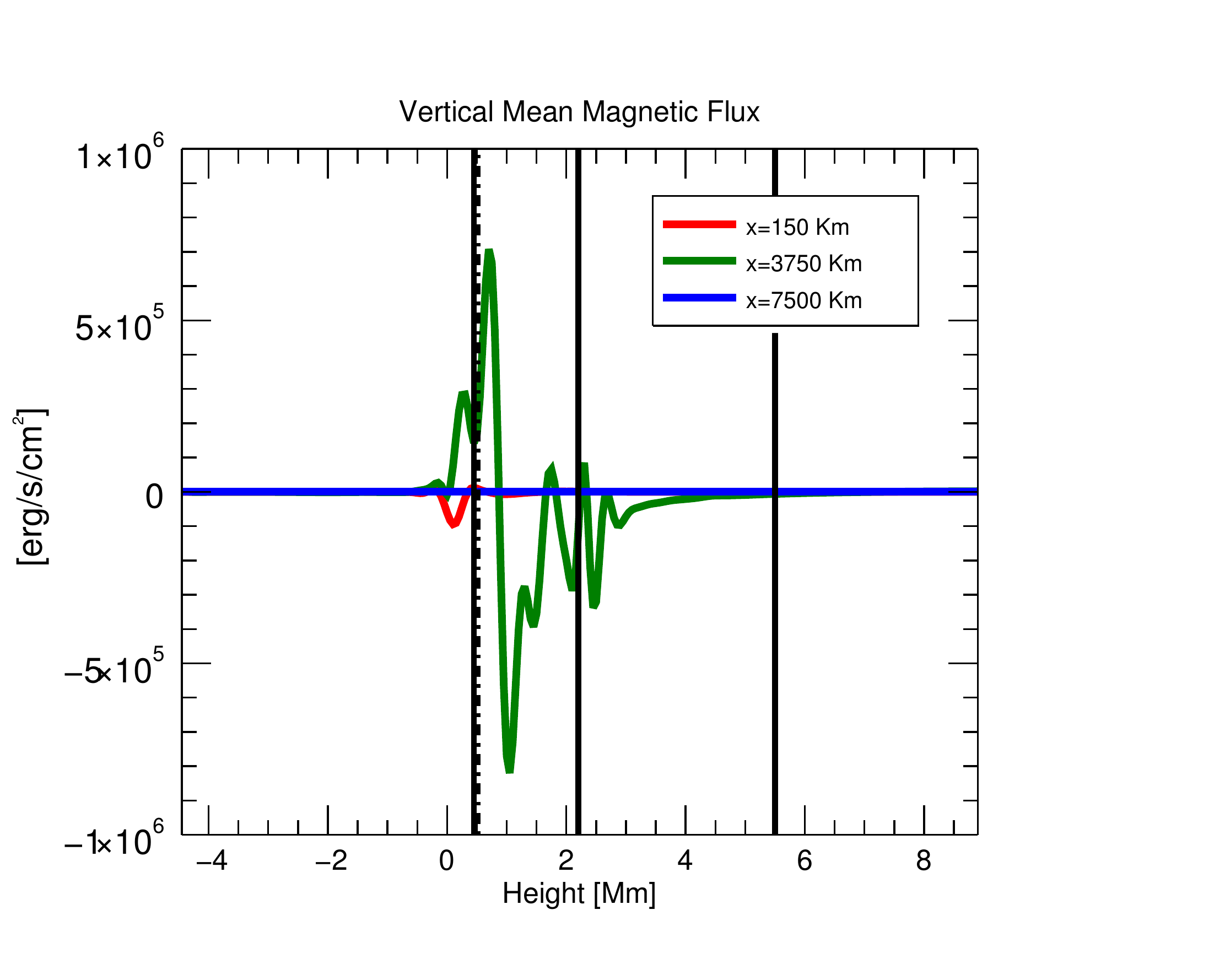}
\caption{Vertical cut of the vertical component of the acoustic (top) and magnetic (bottom) mean fluxes for three different horizontal positions, as indicated in each panel. The vertical black lines are the locations of the cut-off frequency for 200 seconds period waves, $\beta=1$ contours (dashed), transition region and the null point (from left to right). } \label{fig:periodic200fz}
\end{figure}

In Figure \ref{fig:periodic200fz} three different cuts of the mean vertical energy fluxes are shown in more detail. As before, we made cuts at three horizontal locations: inside the flux tubes at $x=150$ km; at the middle of the arcades where the magnetic field is horizontal at $x=3750$ km and at $x=7500$ km where the null point is located. The bottom panel of Fig.  \ref{fig:periodic200fz} clearly shows that all the magnetic energy is concentrated between the photosphere ($z=0$ km) and the transition region, as a consequence of the mode conversion in the inclined magnetic field of the arcades (green line). The sign of the energy flux is negative meaning that waves essentially propagate down, which is explained by the influence of the null point. The magnetic energy flux in our experiment is unable to cross the transition region. 

The upper panel of Fig. \ref{fig:periodic200fz} shows that the acoustic energy in the photosphere propagates upwards in the nearly vertical magnetic concentrations (blue and red lines), and downwards at the locations of the arcade. This acoustic energy flux is significantly larger in the photosphere and in the coronal part, compared to the magnetic flux (the latter is almost zero in these regions). The largest upward acoustic energy flux reach the corona in the nearly vertical flux tubes at the sides of the domain and makes less than 10--20\% of the original energy flux of acoustic waves below the photosphere.

%%%%%%%%%%%%%%%%%%%%%%%%%%%%%%%%%%%%%%%%%%%%%%%
%%%%%%%%%%%%%%%%%%%%%%%%%%%%%%%%%%%%%%%%%%%%%%%

%%%%%%%%%%%%%%%%%%%%%%%%%%%%%%%%%%%%%%%%%%%%%%%
\subsection{Horizontal periodic driver of 300 seconds}
%%%%%%%%%%%%%%%%%%%%%%%%%%%%%%%%%%%%%%%%%%%%%%%

In this run, a horizontal periodic motion with a period of 300 seconds is applied at the lower boundary of the simulation domain at heights  $z=\{-5, -4.65\}$ Mm, same as for the vertical driver. 
\begin{eqnarray}  \label{eq:hor_driver}
\delta v_x &=& V _0 \sin (\omega t) \\
\frac{\delta B_x}{B_0} &=& \frac{k_z}{\omega} V_0 \sin (\omega t)
\end{eqnarray}
where the horizontal wavelength is infinite, so $k_x=0$, and \begin{math} k_{z}=\omega /v_{a} \end{math}.The perturbation in the rest of the variables is null. The value of the magnetic field strength $B_0 = \sqrt{B_{0x}^2 + B_{0z}^2}$ is taken at the boundary. The horizontal velocity amplitude is chosen to be  of $V_0= 0.2$ \ms, see Tab. \ref{tab:setup}. Such driving generates essentially no acoustic waves in vertical magnetic fields where the motion of the driver is completely transverse. This can be seen in some points of the edges of the domain and in the middle, between the two arcades. However there appear acoustic-like waves in almost the rest of the domain, where the magnetic field is more inclined \citep[see][]{Hasan+etal2005, Hasan+Ballegooijen2008}.

%%%%%%%%%%%%%%%%%%%%%%%%%%%%%%%%%%%%%%%%%%%%%%%
\begin{figure} 
\centering
\includegraphics[width=7cm]{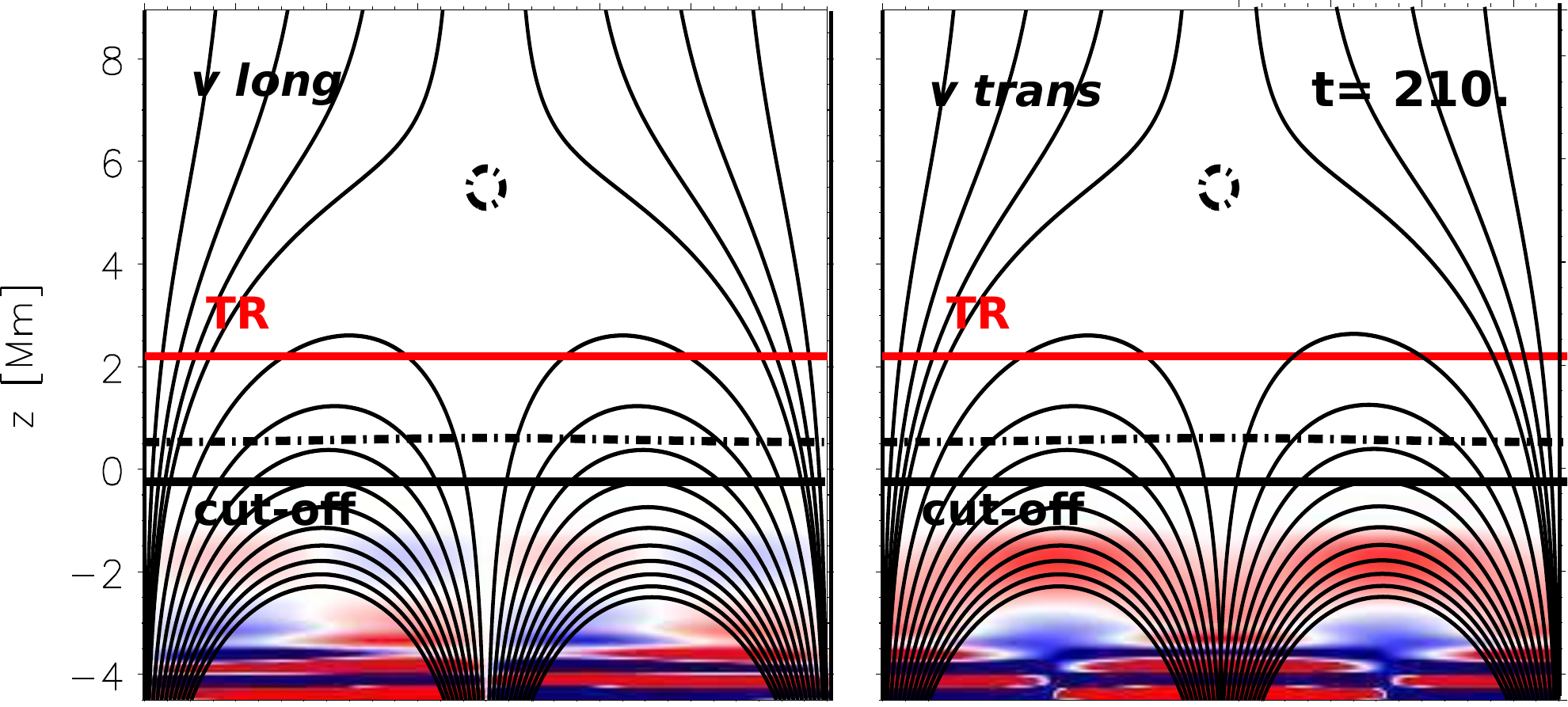}
\includegraphics[width=7cm]{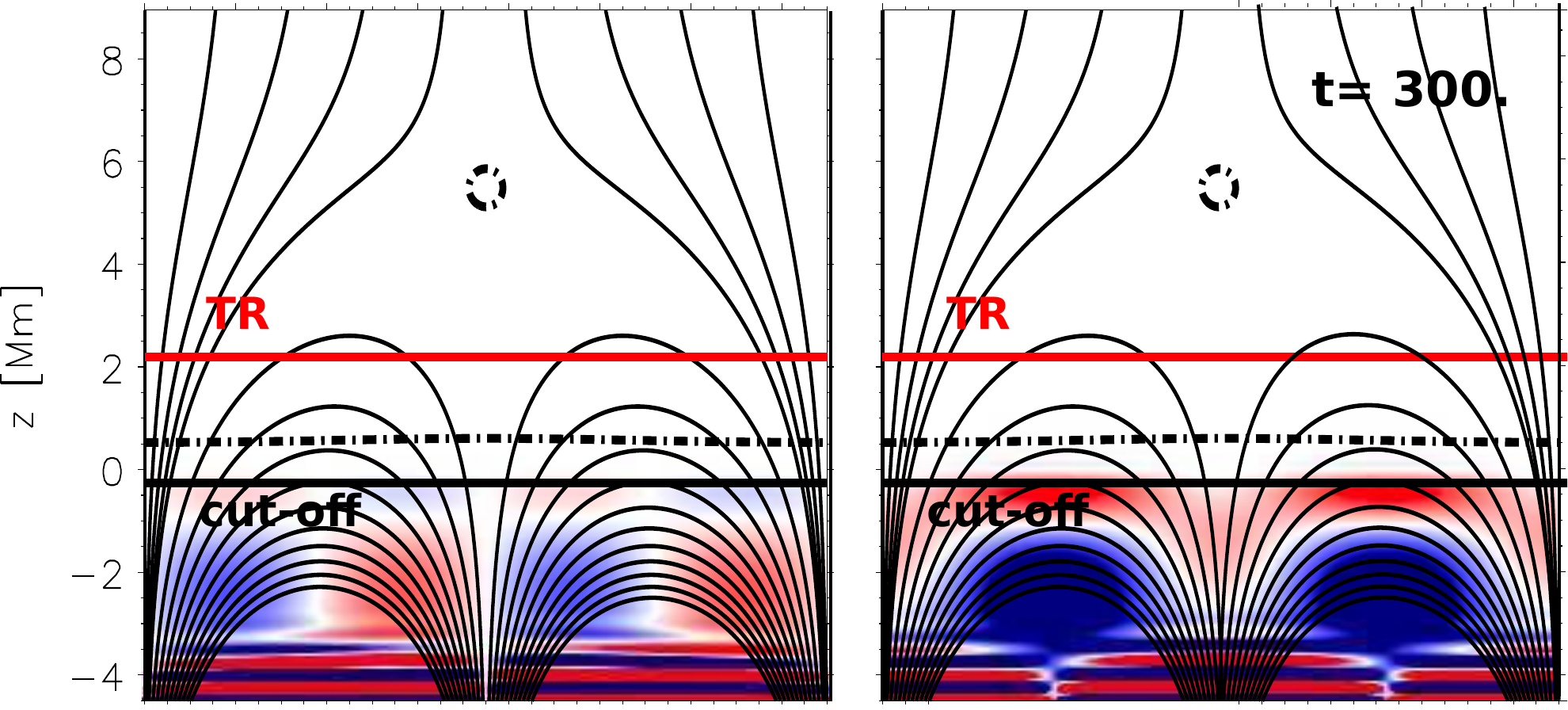}
\includegraphics[width=7cm]{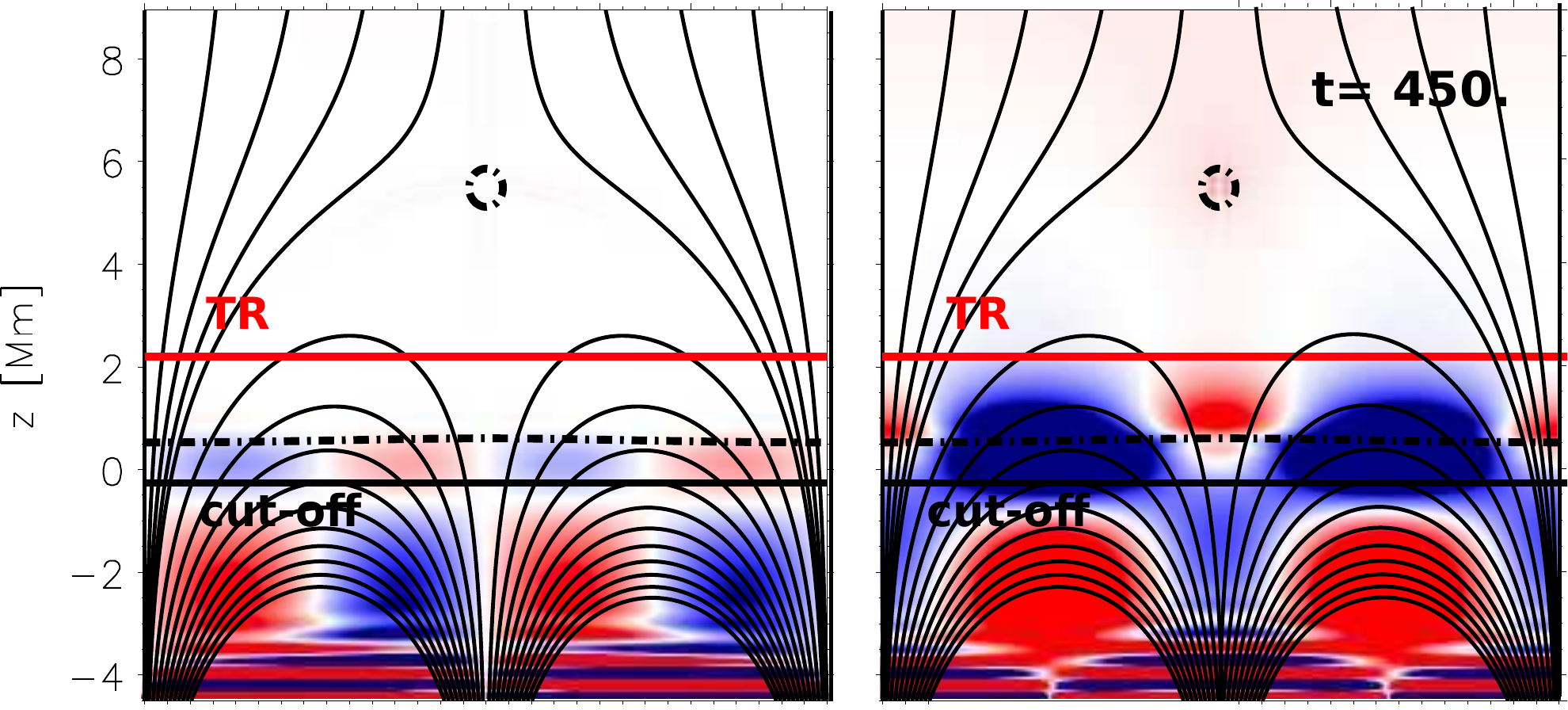}
\includegraphics[width=7cm]{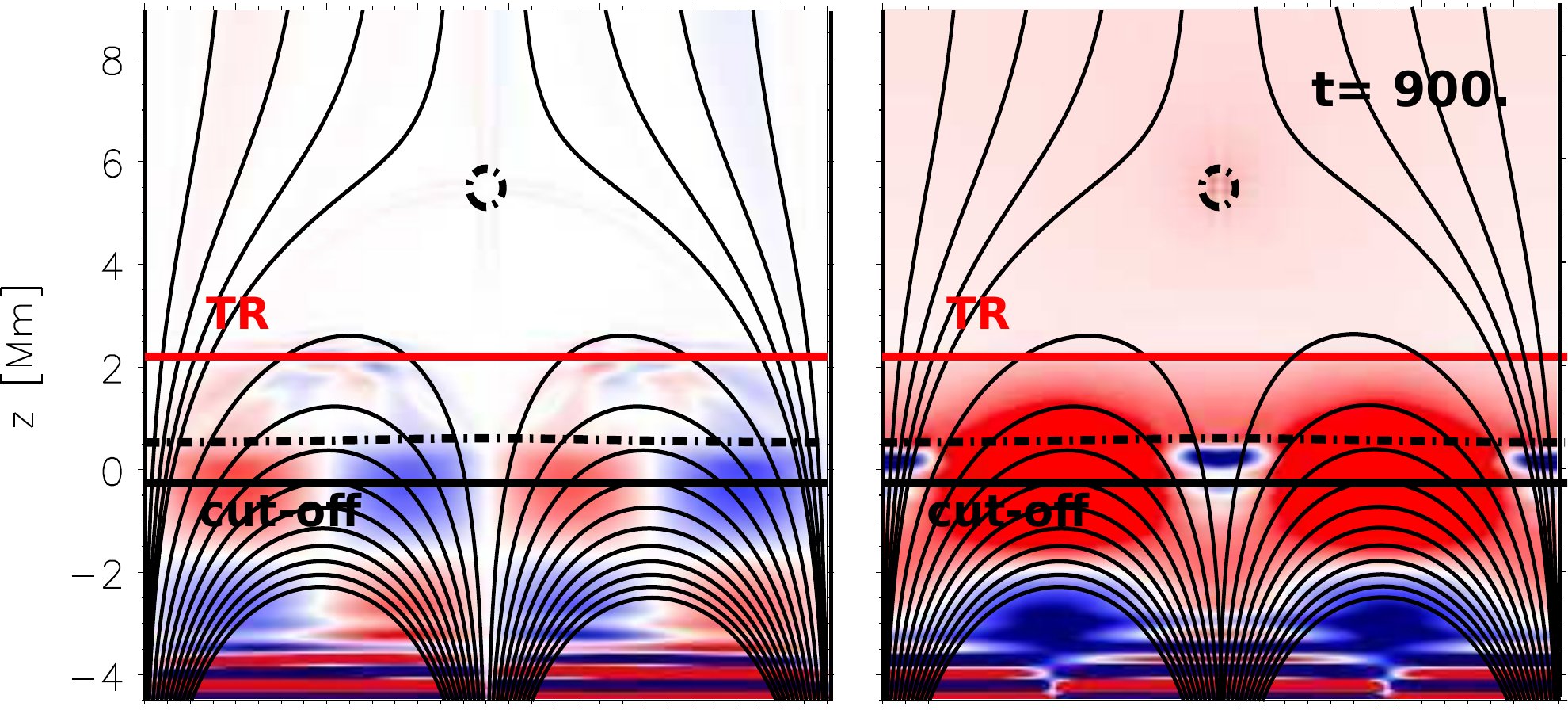}
\includegraphics[width=7cm]{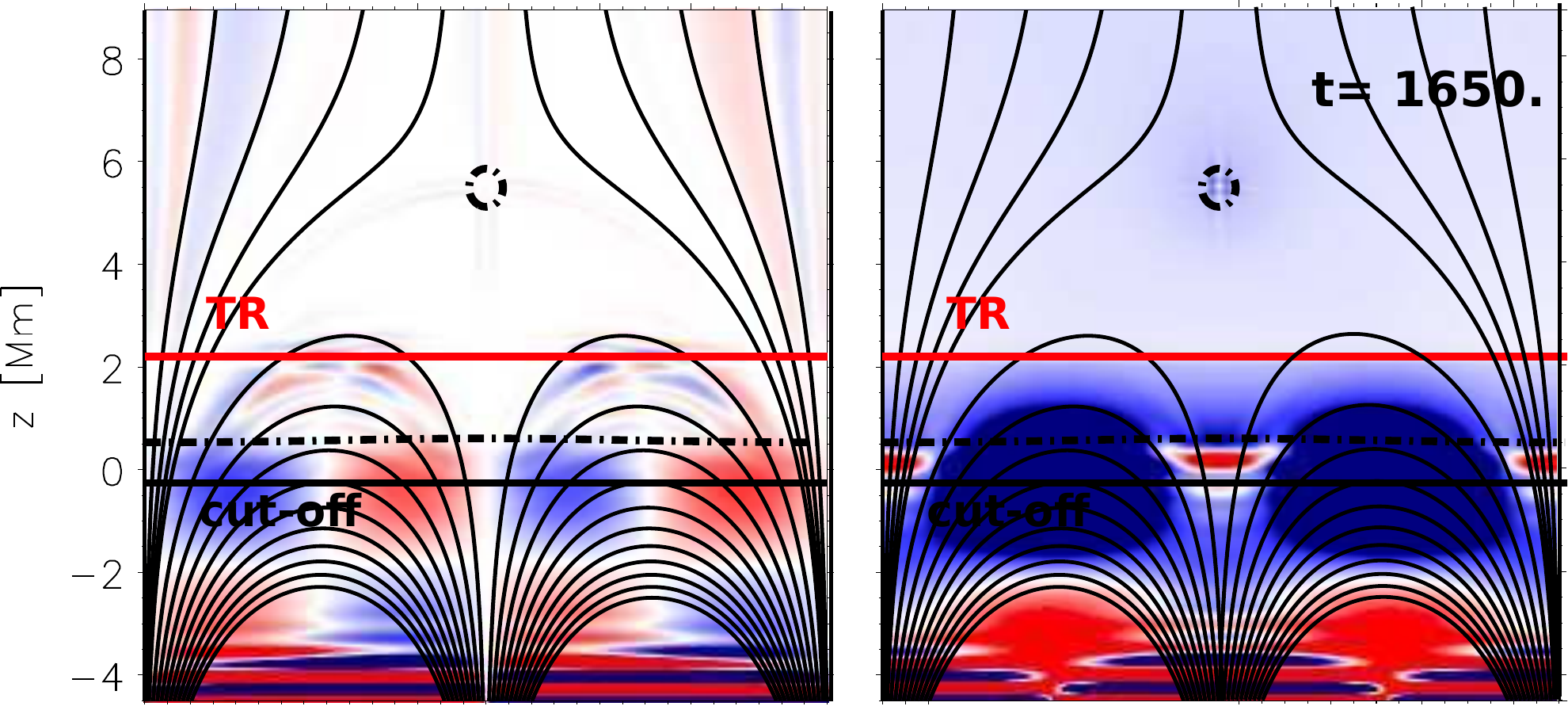}
\includegraphics[width=7cm]{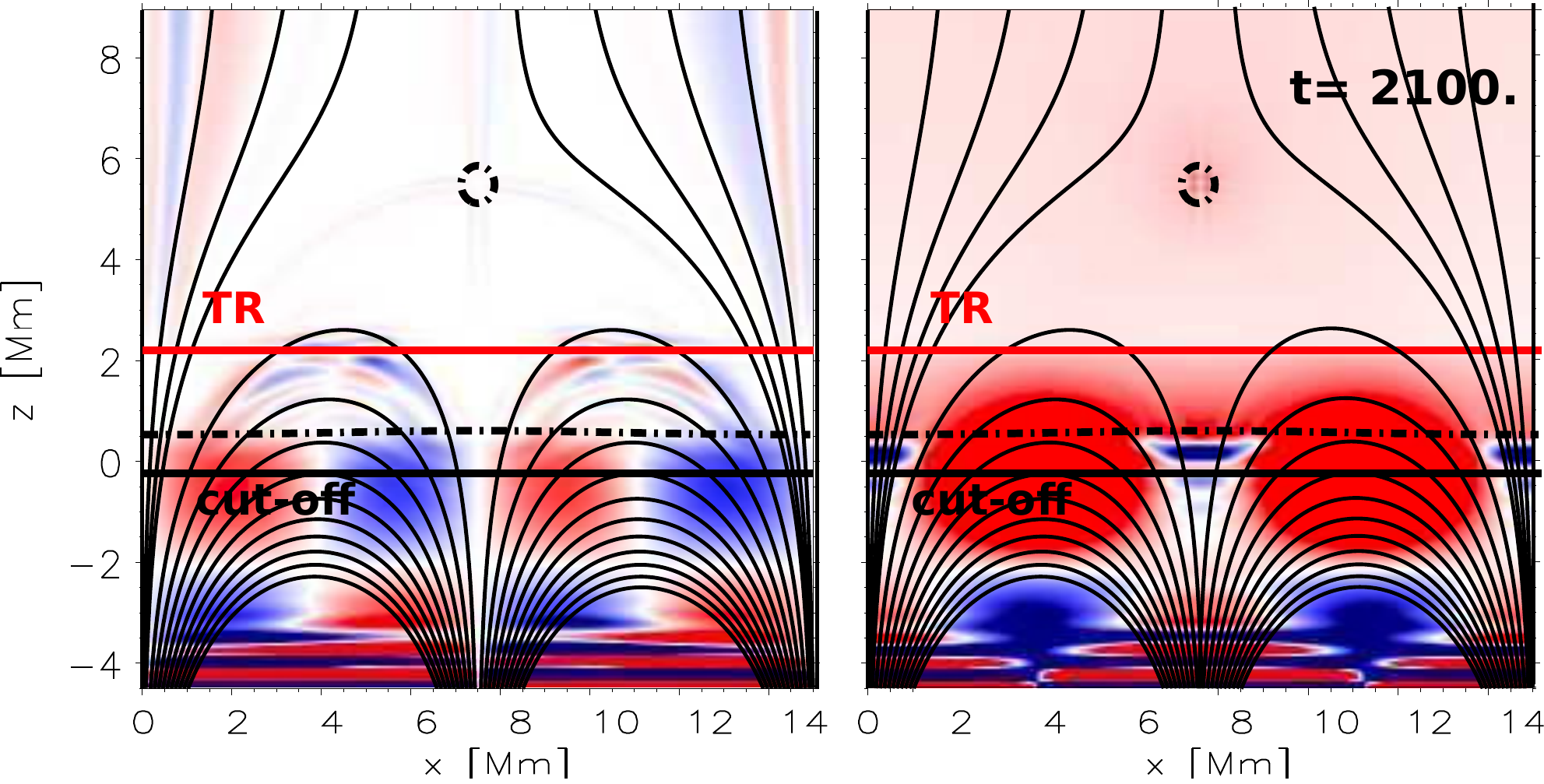}
\includegraphics[width=7.0cm]{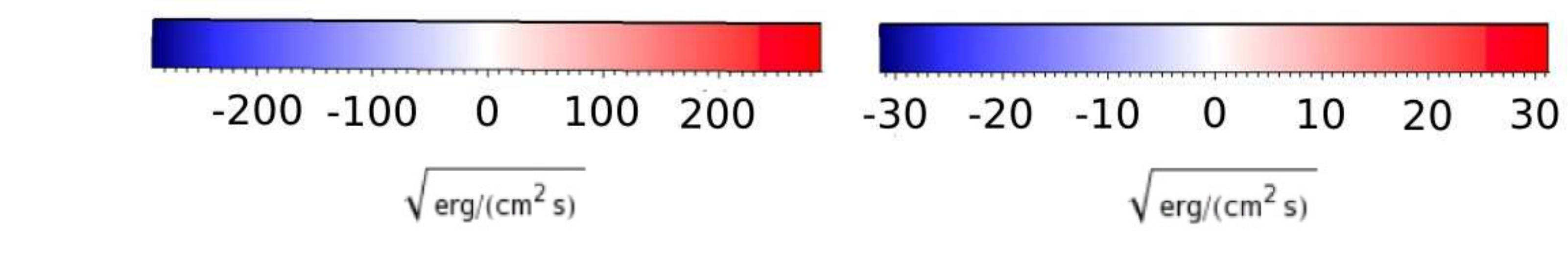} 
\caption{Same as Fig. \ref{fig:vlvt} but for the simulation run with a horizontal periodic driving of 300 s. The black solid line is the layer where the cut-off period is 300 s. The movie is vailable online.} \label{fig:vlvt_horiz}
\end{figure}
%%%%%%%%%%%%%%%%%%%%%%%%%%%%%%%%%%%%%%%%%%%%%%%

%%%%%%%%%%%%%%%%%%%%%%%%%%%%%%%%%%%%%%%%%%%%%%%
\subsubsection{Wave propagation}
%%%%%%%%%%%%%%%%%%%%%%%%%%%%%%%%%%%%%%%%%%%%%%%

Figure \ref{fig:vlvt_horiz} shows the propagation of waves generated by the horizontal periodic driver. The driver excites magnetic slow waves seen as a small wavelength pattern at the bottom of the domain at all times.  The pattern due to these waves disappears after some height around $-3$ Mm since these small-wavelength perturbations are affected by the numerical diffusion effects and are not fully resolved. At the locations with inclined field, fast acoustic waves are also excited. These can be appreciated as a large-scale disturbance propagating from the bottom boundary upwards at the two upper panes and at later times as well. These fast waves reach the equipartition layer at $t=300$ s, and they get partially converted and transmitted. The perturbation seen in longitudinal velocity (left panels) remains almost completely below the equipartition layer at all posterior time moments. The only exception are the vertical flux tubes at the domain sides where again, similar to the case of the vertical driving, some perturbation propagates upwards through the transition region to the corona. 

Perturbations seen in the transverse velocity (right panels) above the equipartition layer and below the transition region at later times ($t>450$ s) correspond to fast magnetic waves. Neither these fast magnetic waves, nor slow acoustic waves (small-scale perturbations between the layer where $\beta=1$ and the transition region at the left panels), pass through the transition region with large amplitudes, and their energy remains essentially below it. The amplitudes of waves reaching the corona are smaller than in the case of the vertical driver. 

Unlike the longitudinal slow waves in vertical flux tubes, fast magnetic transverse waves reach the corona in the whole domain. When passing through the null point, they are partially converted into acoustic-like waves in the equipartition layer around it. The simulation reaches a stationary state after about $t=1200$ seconds of the simulation. 

%%%%%%%%%%%%%%%%%%%%%%%%%%%%%%%%%%%%%%%%%%%%%%%
\begin{figure}
\centering
\includegraphics[width=9cm]{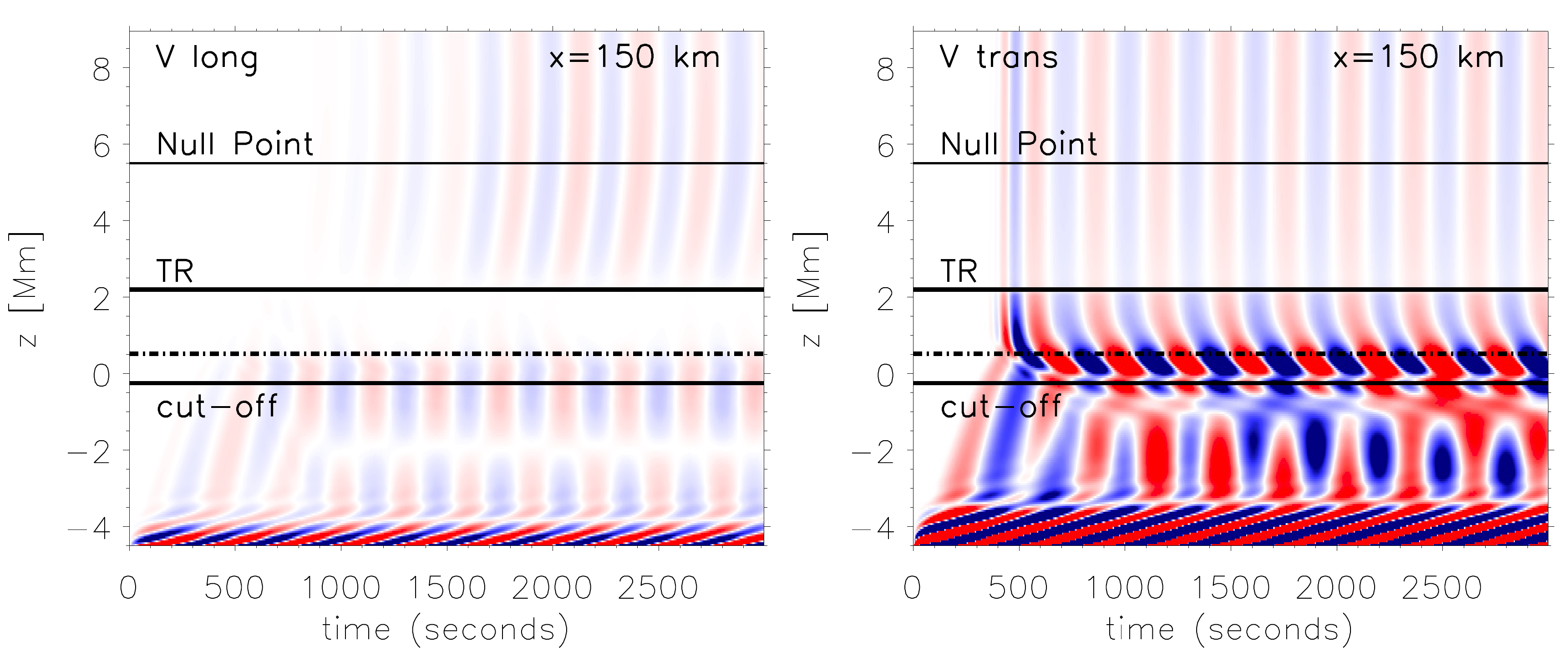}
\includegraphics[width=9cm]{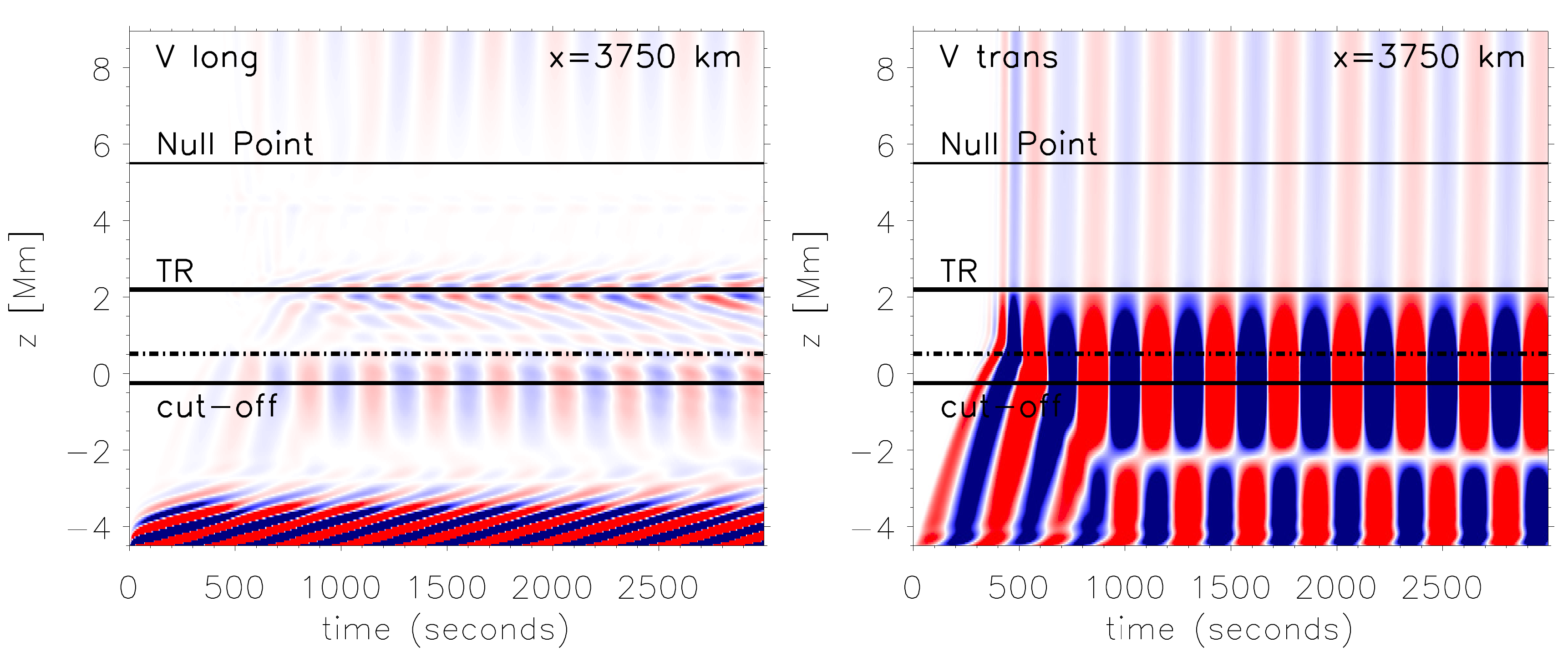}
\includegraphics[width=9cm]{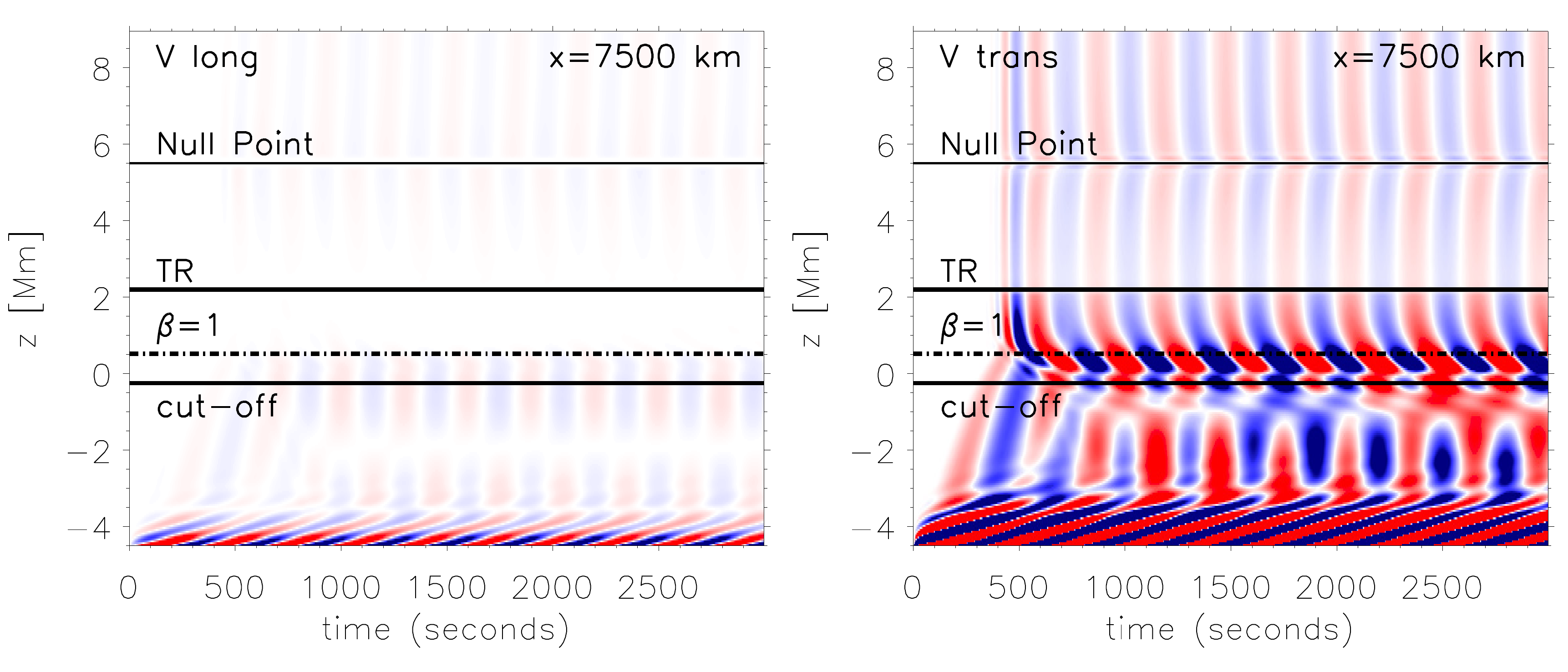}
\caption{Same as Fig. \ref{fig:periodic200td} but for the simulation run with a horizontal periodic driving of 300 s.} \label{fig:periodic300td}
\end{figure}
%%%%%%%%%%%%%%%%%%%%%%%%%%%%%%%%%%%%%%%%%%%%%%%

Figure \ref{fig:periodic300td} shows the time-height diagrams of $\sqrt{\rho_0 c_{s}}v_{\rm long}$ (left) and $\sqrt{\rho_0 v_{a}}v_{\rm trans}$ (right) at three different horizontal positions of the domain. The top and bottom rows of this figure (horizontal positions $x=150$ km and $x=7500$ km) show waves propagating in the nearly vertical magnetic field. One can appreciate how the behaviour of these waves changes at the equipartition layer (dashed-dotted line) and at the transition region. The influence of the cut-off layer is not so clear in these diagrams. The transverse velocity propagation at these two positions of the domain is very similar. The mode conversion results in an increase of the transverse wave component as the waves cross the equipartition layer in the downward direction. 
The ridges on the diagram show the opposite inclination indicating downward propagation produced by the reflection at the transition region. 
In the corona, the ridges are vertical, indicating that waves propagate with Alfv\'en speed, so their nature is magnetic. As these fast waves cross the null point, they are converted at the equipartition layer around it (bottom right panel of Fig. \ref{fig:periodic300td}). The propagation speed of the waves is altered, which results in a shift of the position of the ridges at heights around the null point. 

The longitudinal velocity behaviour at $x=150$ km and at $x=7500$ km is quite similar, except of the amplitude of the waves. Some weak longitudinal slow waves appear above the transition region at $x=150$ km, generated by horizontally propagating perturbations in the corona. 

The middle panel of Fig. \ref{fig:periodic300td} shows the wave propagation at the location of the inclined field of the arcade. There, the transverse waves propagate undisturbed up to the transition region where they are reflected. Conversion from fast acoustic into fast magnetic waves occur at the equipartition level as can be seen from the smooth transition in the inclination of the ridges at that location. 
Some weak acoustic slow waves are also generated there (middle left panel). These waves propagate up to the transition region and are reflected there and form an interference pattern around 2 Mm height. 

%%%%%%%%%%%%%%%%%%%%%%%%%%%%%%%%%%%%%%%%%%%%%%%
\subsubsection{Energy fluxes}
%%%%%%%%%%%%%%%%%%%%%%%%%%%%%%%%%%%%%%%%%%%%%%%

Same as for the simulations with a vertical driver we calculated the mean acoustic and magnetic fluxes, see Eqs. \ref{eq:fluxes}, averaging over the stationary state of the simulations, from $t= 1200$ s to $t= 3000$ s (6 periods of 300 s waves). 

%%%%%%%%%%%%%%%%%%%%%%%%%%%%%%%%%%%%%%%%%%%%%%%
\begin{figure}
\centering
\includegraphics[width=9.0cm]{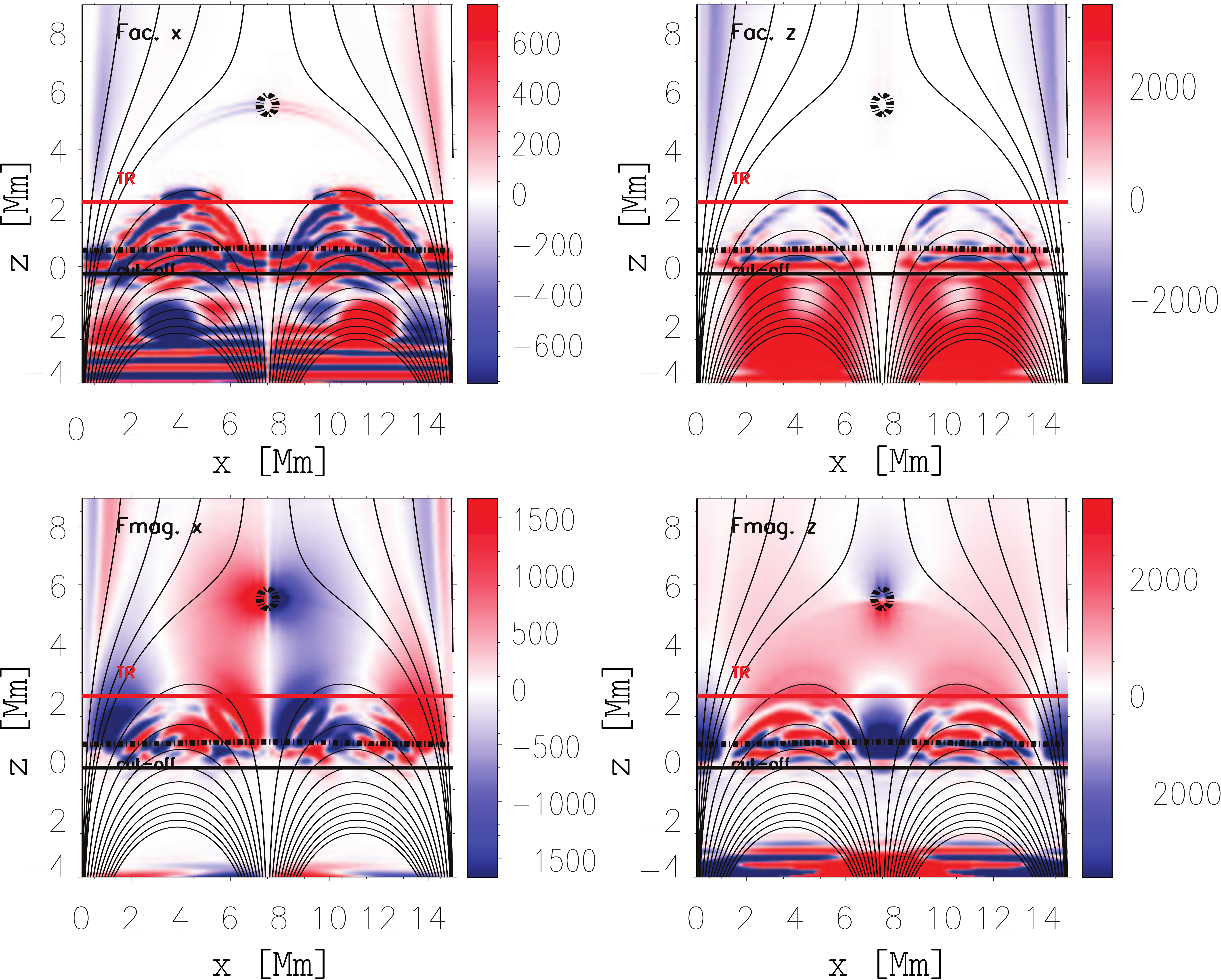}

\caption{Same as Fig. \ref{fig:periodic200flux} but for the simulation run with a horizontal periodic driving of 300 s. The units of the bar are [erg/s/cm$^2$].} \label{fig:periodic300flux}
\end{figure}
%%%%%%%%%%%%%%%%%%%%%%%%%%%%%%%%%%%%%%%%%%%%%%%

%%%%%%%%%%%%%%%%%%%%%%%%%%%%%%%%%%%%%%%%%%%%%%%
\begin{figure}
\centering
\includegraphics[width=10cm]{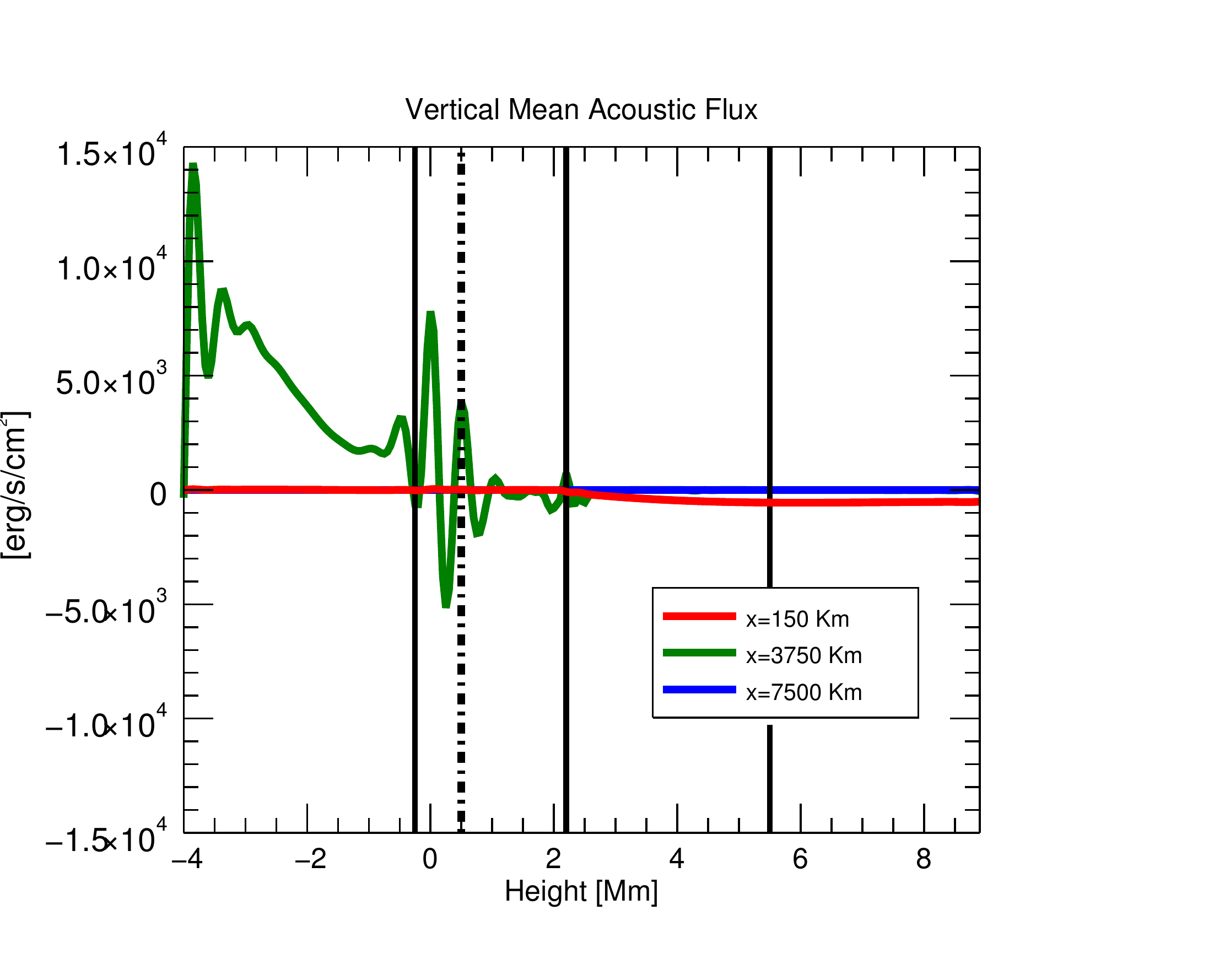}
\includegraphics[width=10cm]{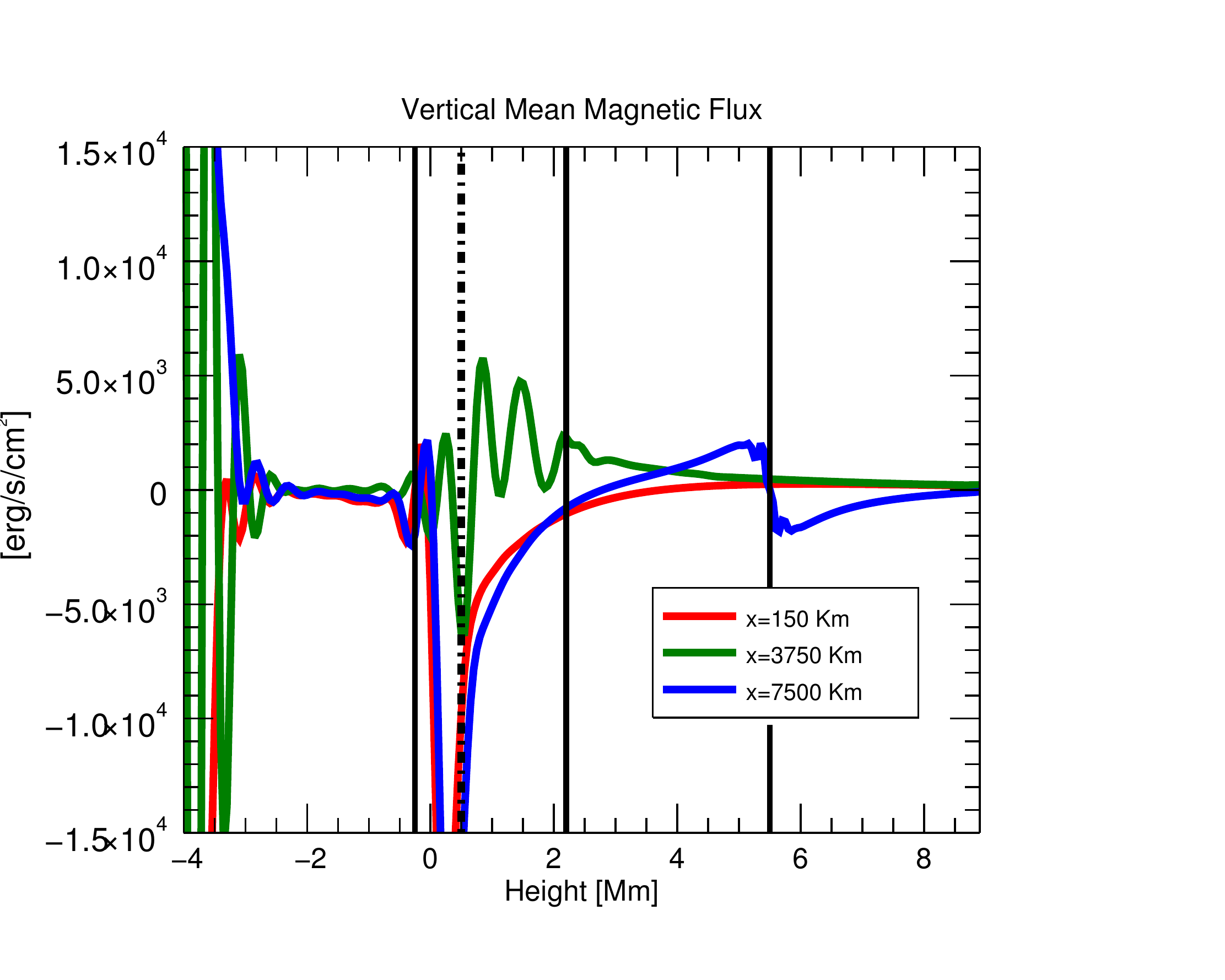}
\caption{Same as Fig. \ref{fig:periodic200fz} but for the simulation run with a horizontal periodic driving of 300 s.} \label{fig:periodic300fz}
\end{figure}
%%%%%%%%%%%%%%%%%%%%%%%%%%%%%%%%%%%%%%%%%%%%%%%

In the case of the horizontal driver the energy fluxes are very small, see Figure \ref{fig:periodic300flux}. This figure shows the spatial distribution of the mean energy fluxes. The upward acoustic energy flux is essentially concentrated below the arcades at heights below the transition region (upper right panel) and does not reach the corona. Curiously, the average upward acoustic flux is negative inside the vertical tubes on the sides, opposite to the case of the vertical driving. This might be related to the fact that in the simulations with the horizontal driver these waves are excited by horizontally propagating disturbances in the corona after the waves are refracted downward around the null point. The vertical magnetic energy flux (bottom right panel) is mostly positive and is much higher above the photosphere than below.

Figure \ref{fig:periodic300fz} shows the vertical magnetic and acoustic fluxes at three different horizontal locations. This figure reveals that the largest acoustic energy flux in this simulation run is concentrated below the arcades (green line at the upper panel). The magnetic flux is very small except at heights around the equipartition region. The values of the flux are much lower than in the vertical driving case, despite the its larger driving amplitude.  

%%%%%%%%%%%%%%%%%%%%%%%%%%%%%%%%%%%%%%%%%%%%%%%
\subsection{Instantaneous pressure pulse}
%%%%%%%%%%%%%%%%%%%%%%%%%%%%%%%%%%%%%%%%%%%%%%%

In this run we exerted a localized instantaneous force at the base of the arcade, in the middle of the domain, that consists of a pressure pulse of a gaussian shape, as follows: 
\begin{eqnarray} \label{eq:pulse_p}
\frac{\delta p_{1}}{p_0} = A \gamma \exp \Big [-\Big( \frac{(x-x_{0})^{2}}{2\sigma_{x}^{2}} + \frac{(z-z_{0})^{2}}{2\sigma_{z}^{2}} \Big) \Big] \\
\frac{\delta \rho_{1}}{\rho_0} = A \exp \Big [-\Big( \frac{(x-x_{0})^{2}}{2\sigma_{x}^{2}} + \frac{(z-z_{0})^{2}}{2\sigma_{z}^{2}} \Big) \Big]
\end{eqnarray}
where A=10$^{-5}$ is the relative amplitude given to the pulse, $[x_0,z_0]=[7500,-3500 ]$ km are the coordinates at which the pulse is located and, $\sigma_{x}=1500$ km and $\sigma_{z}=1000$ km are the widths of the two-dimensional Gaussian profile, see Tab. \ref{tab:setup}. Such a pulse generates a superposition of waves in a broad range of frequencies.

Unlike the simulations with the vertical and horizontal driving, we locate a PML boundary condition at both, top and bottom, boundaries of our simulation domain. 

%%%%%%%%%%%%%%%%%%%%%%%%%%%%%%%%%%%%%%%%%%%%%%%
\begin{figure} \label{fig:vlvt_pulse}
\centering
\includegraphics[width=7cm]{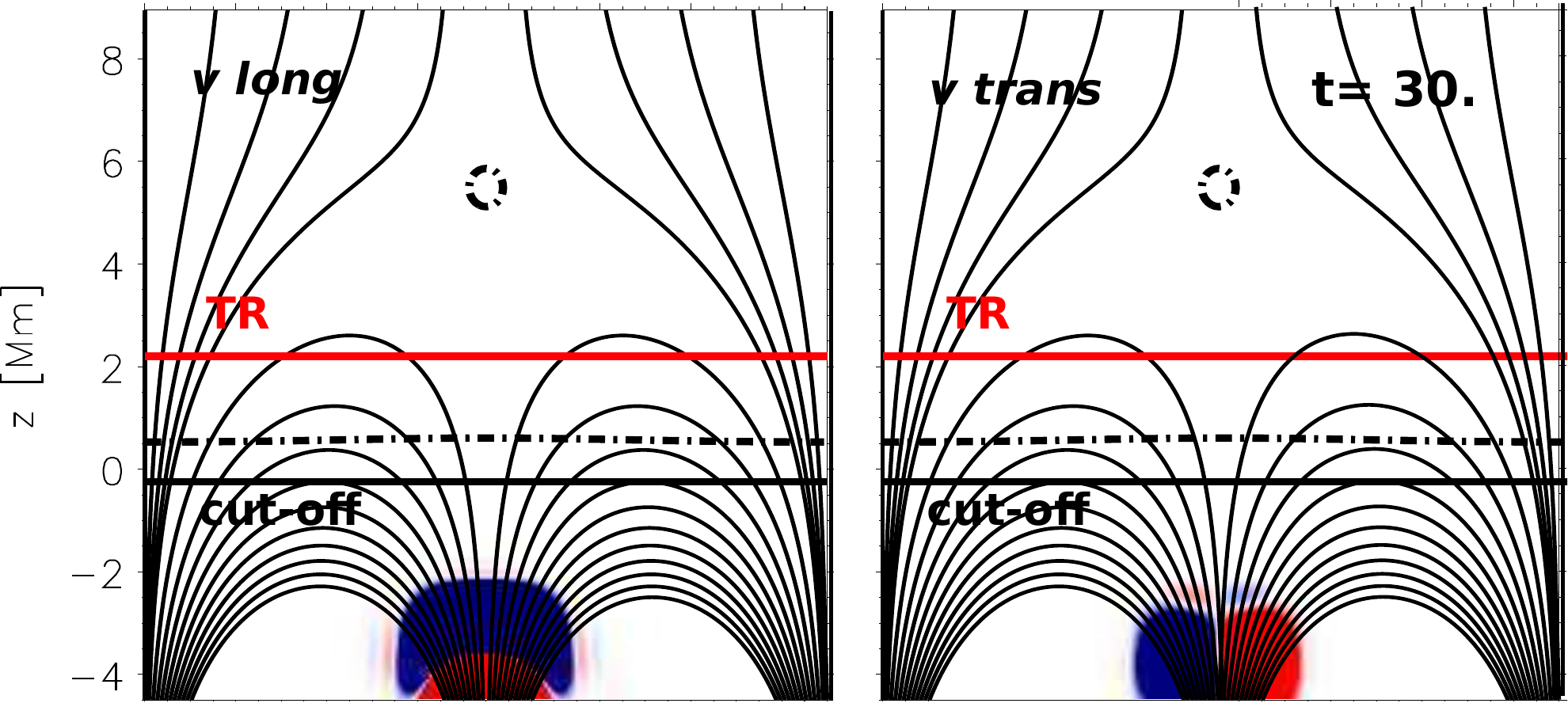} 
\includegraphics[width=7cm]{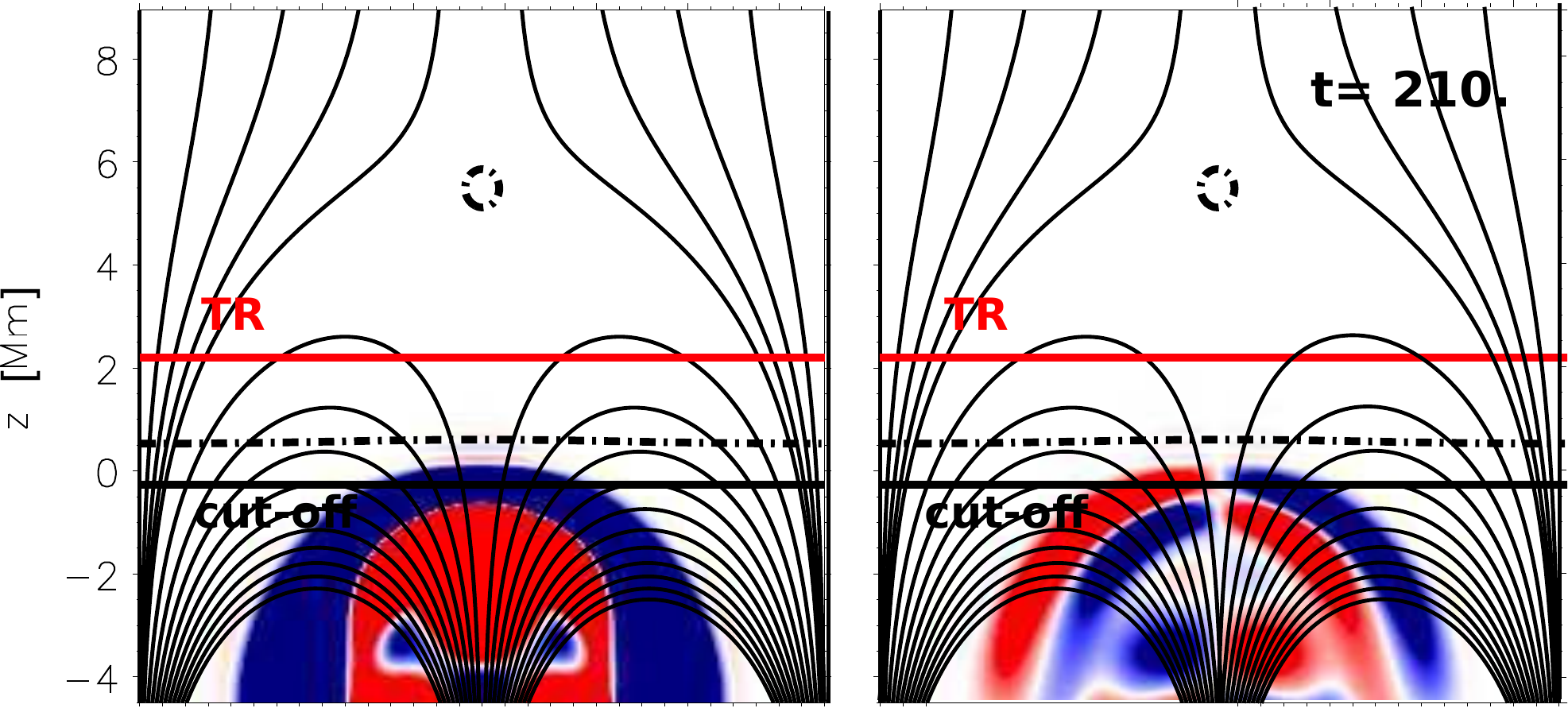} 
\includegraphics[width=7cm]{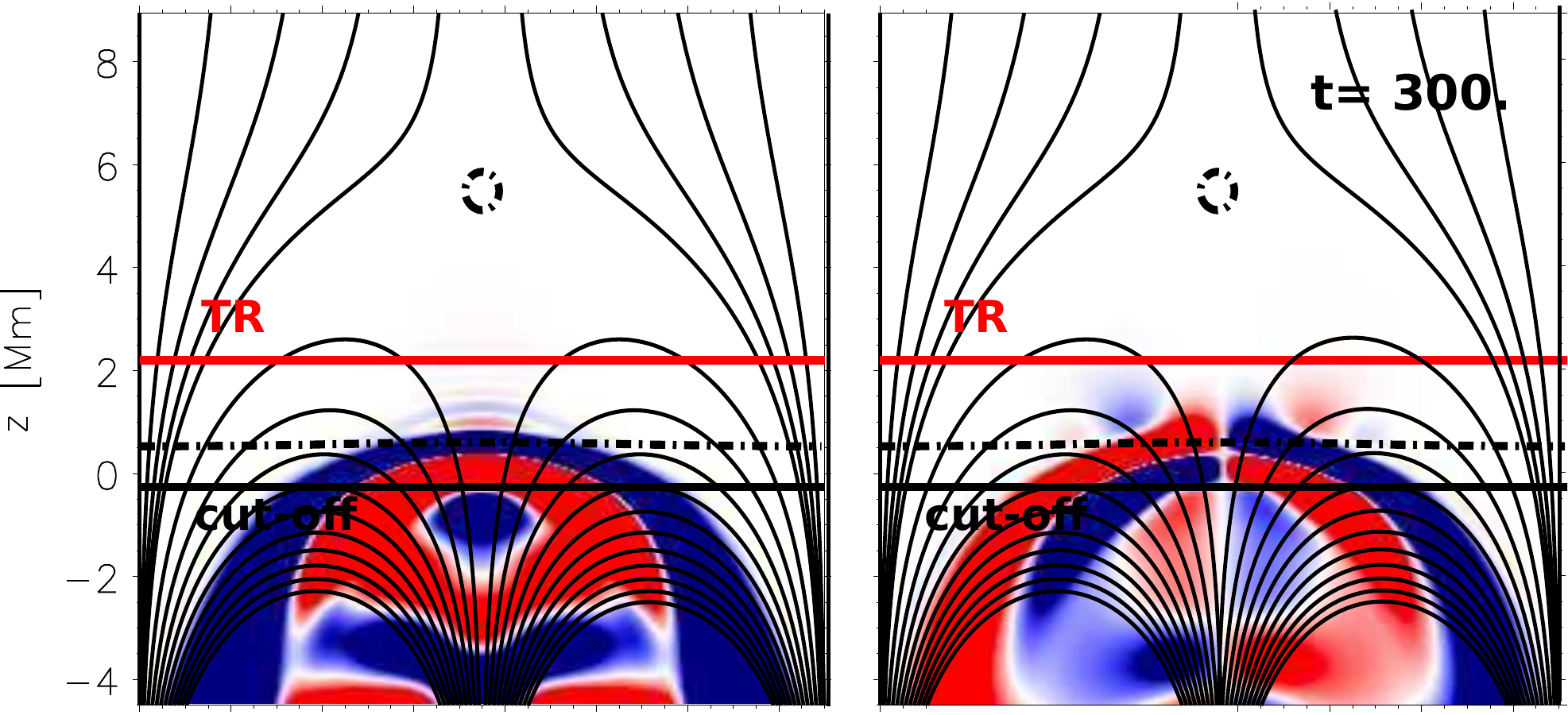}  
\includegraphics[width=7cm]{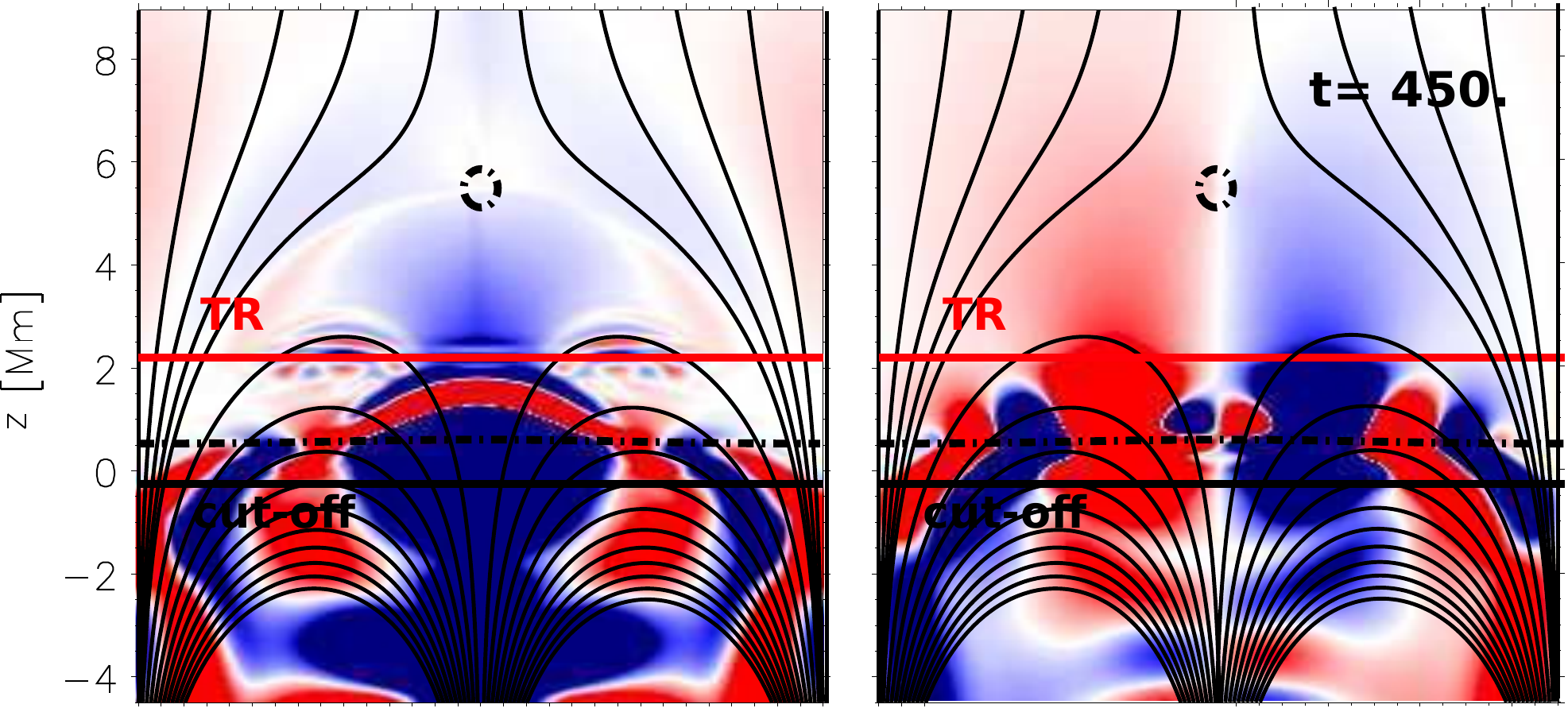} 
\includegraphics[width=7cm]{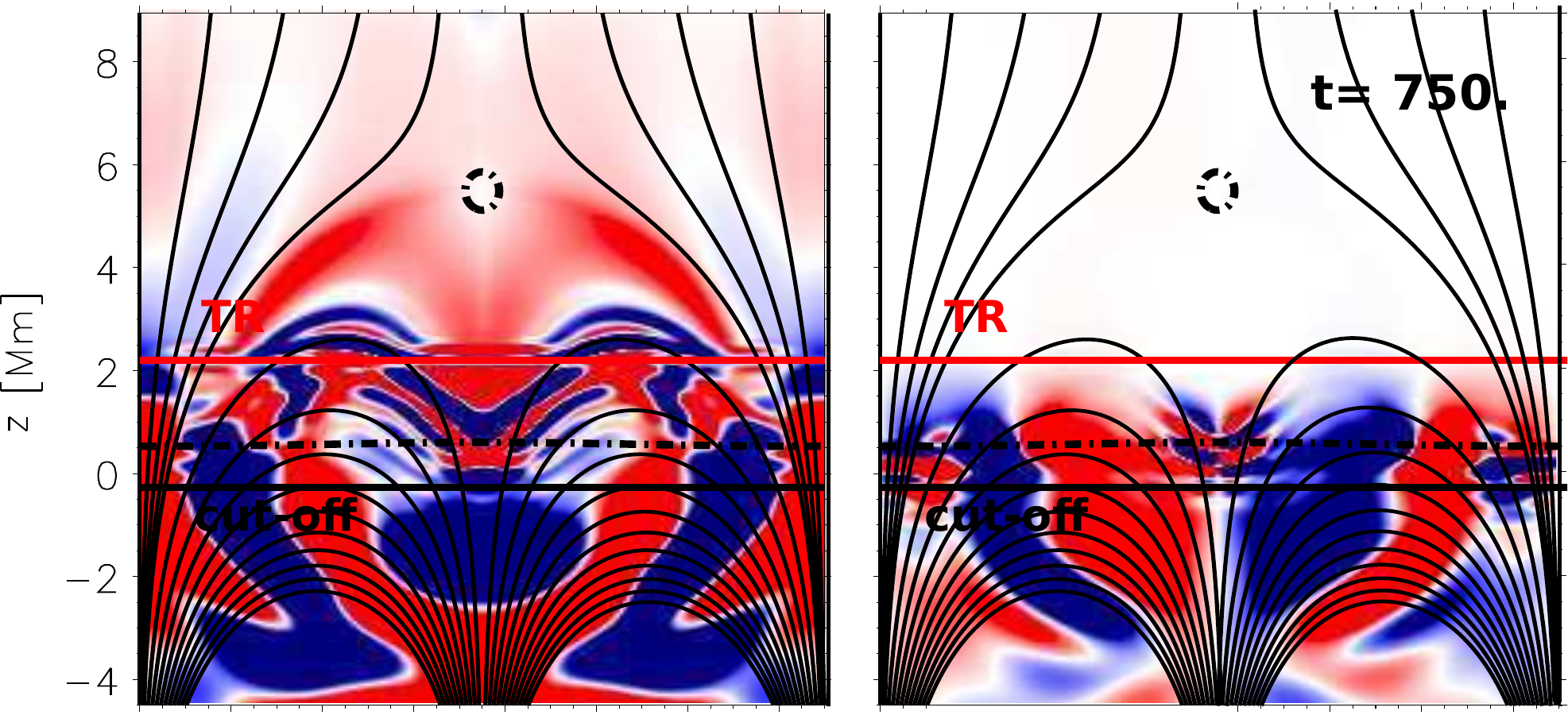}
\includegraphics[width=7cm]{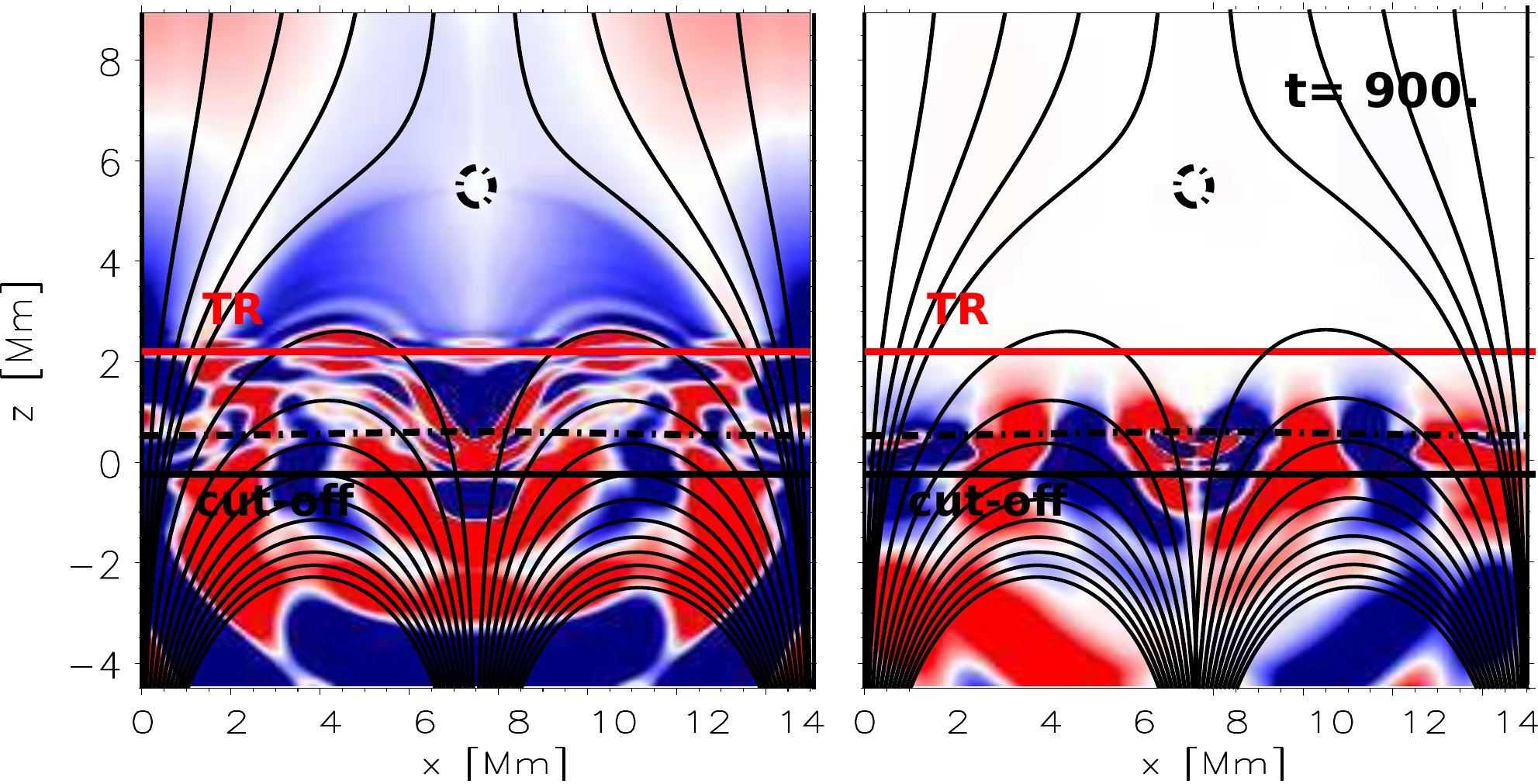}
\includegraphics[width=7.0cm]{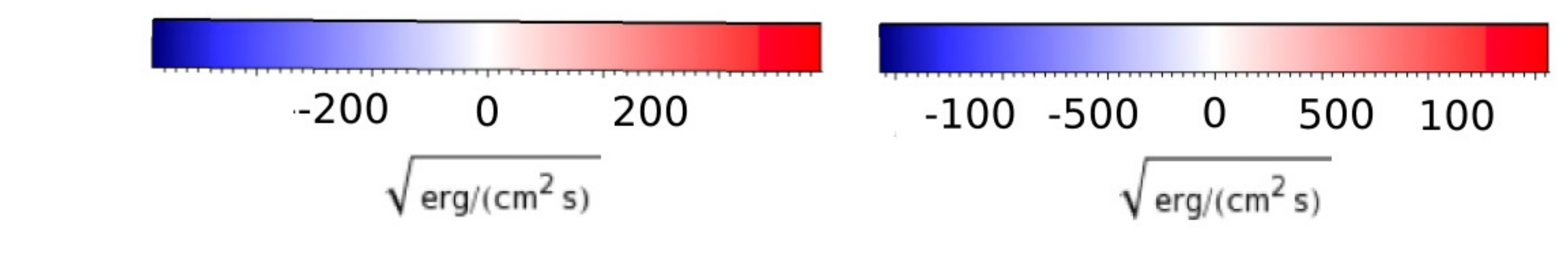} 
\caption{Same as Fig. \ref{fig:vlvt} but for the simulations with an instantaneous pressure pulse. The movie of the wave propagation is available in the online version of the paper. } \label{fig:vlvt_pulse}
\end{figure}
%%%%%%%%%%%%%%%%%%%%%%%%%%%%%%%%%%%%%%%%%%%%%%%

%%%%%%%%%%%%%%%%%%%%%%%%%%%%%%%%%%%%%%%%%%%%%%%
\subsubsection{Wave propagation}
%%%%%%%%%%%%%%%%%%%%%%%%%%%%%%%%%%%%%%%%%%%%%%%

The time evolution of the scaled longitudinal and transverse velocity components of the pulse simulation are given in Figure \ref{fig:vlvt_pulse}. The pulse starts expanding in all the directions (two upper panels). Like in the other runs, the waves get partially converted and transmitted when reaching the equipartition layer around $t=300$ s. The acoustic fast component of the pulse is partially converted and transmitted into magnetic fast and acoustic slow waves, respectively.

The acoustic slow waves continue propagating upward and reach the transition region before $t=450$ s. Part of these waves is  transmitted into the corona and they reach the null point, where they become trapped and are sent again in all directions, except those waves inside the vertical flux tubes that continue propagating to upper layers along the field lines (left panels of Fig. \ref{fig:vlvt_pulse}). 

Acoustic waves reflected from the transition region reach equipartition layer again on their downward propagation, and they suffer another partial transmission. Fast acoustic waves produced after this secondary transmission propagate down back to the lower boundary of the domain, where they get refracted due to the gradients of the acoustic speed and are finally reflected up around $-4$ Mm at the lower turning point. Therefore, these waves start propagating upwards again (see left and right panels of Fig. \ref{fig:vlvt_pulse} at $t=750$ s at the lower part of the domain) suffering the same phenomena as before. Thus, despite our driver is not periodic, the atmosphere keeps oscillating reaching a nearly stationary regime without a significant decrease of the wave power during all 3000 s of the simulation time (only waves reaching the upper and lower PML boundaries are removed from the system). The null point together with the transition region and the lower turning point below the photosphere act as a re-feeding of the atmosphere, and the waves are trapped between these layers. 

The variations of the transverse velocity (right panels of Fig. \ref{fig:vlvt_pulse}) reveal that fast magnetic waves are produced after the mode conversion around 300 s, and then reach the transition region and are partially transmitted to the corona before $t=450$ s. These fast magnetic waves get refracted and reflected due to the gradients of the Alfv\'en speed, and cross again the equipartition layer on their way back. On their way back, the fast magnetic waves are partially transmitted into slow magnetic waves with small wavelengths, that propagate downwards and their energy gets lost due to the numerical diffusion. At the moments posterior to $t=750$ s, no perturbations in the transverse velocity can be appreciated. 

%%%%%%%%%%%%%%%%%%%%%%%%%%%%%%%%%%%%%%%%%%%%%%%
\begin{figure}
\centering
\includegraphics[width=9cm]{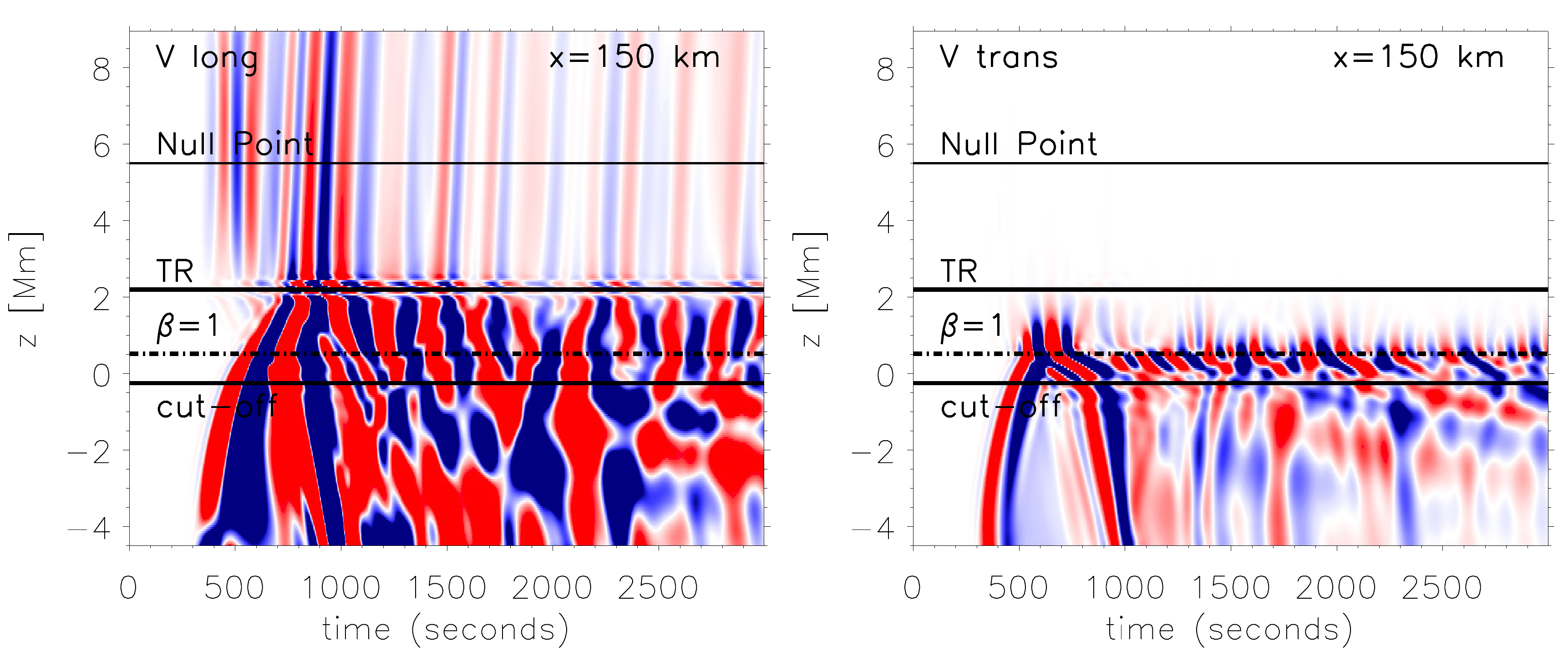}
\includegraphics[width=9cm]{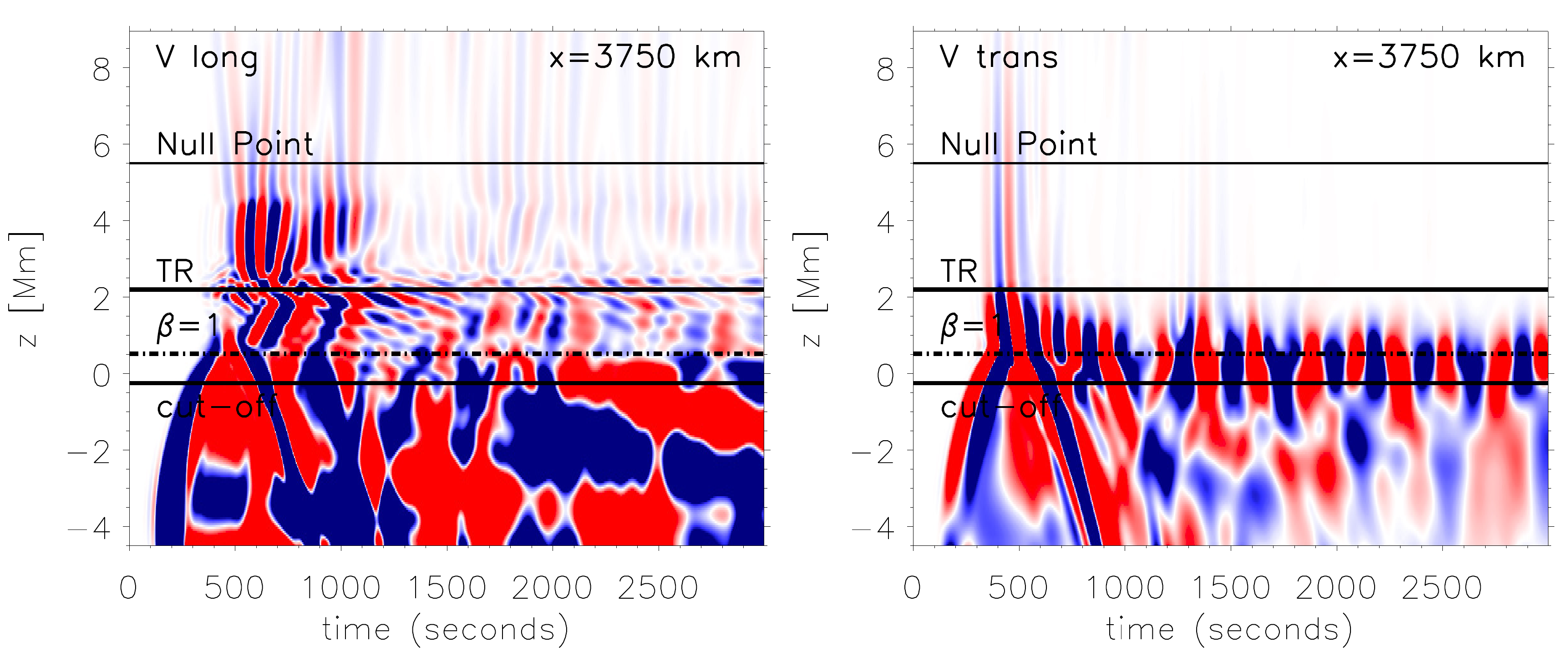}
\includegraphics[width=9cm]{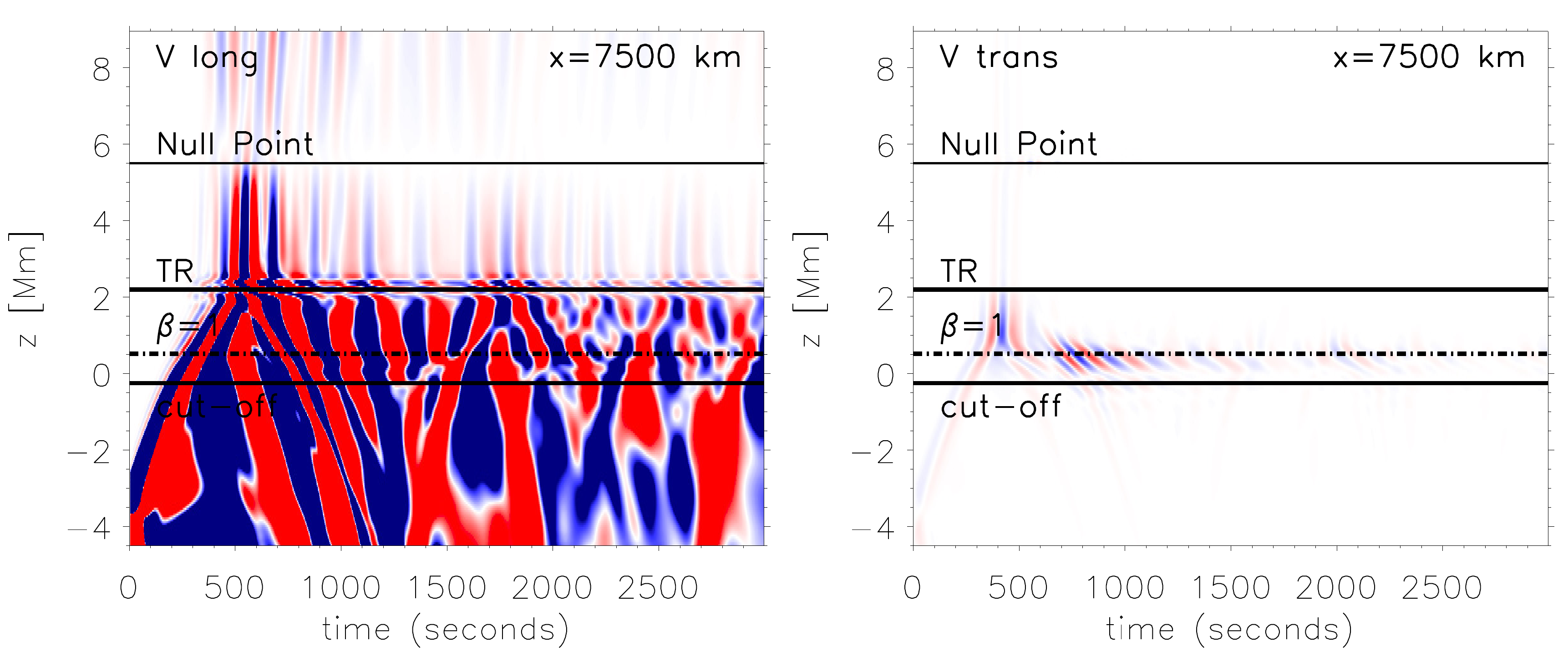}
\caption{Same as Fig. \ref{fig:periodic200td} but for the simulations with an instantaneous pressure pulse.} \label{fig:pulse_td}
\end{figure}
%%%%%%%%%%%%%%%%%%%%%%%%%%%%%%%%%%%%%%%%%%%%%%%

Same as for the other cases, Figure \ref{fig:pulse_td} shows the time-height diagrams of the scaled longitudinal and transverse velocities at three representative horizontal locations of the domain. It is clearly seen from this figure that, indeed,  the atmosphere keeps oscillating during all the simulation time, but the stationary regime is not reached since the variations in time are not strictly periodic, as was for the case of harmonic driving. There are few strong initial wavefronts that can be appreciated in the corona for $t<1000$ s. Then, the amplitude of the coronal oscillations becomes gradually smaller. The dominant waves propagating in the corona are acoustic-like slow waves, and those with largest amplitudes are observed inside the vertical flux tubes (upper left panel for $x=150$ km), consistent with the simulations of periodic driving. No appreciable fast magnetic waves are present after about $t=1000$ of the simulation time, due to the reasons explained above. Fast acoustic waves propagating upwards from the lower turning point can also be appreciated at the left panels of Fig. \ref{fig:pulse_td} at the bottom part of the domain. 

Since the pulse simulation does not reach the stationary regime, we are not able to calculate and analyze the average wave fluxes, as was done for the periodic driver simulations. 

%%%%%%%%%%%%%%%%%%%%%%%%%%%%%%%%%%%%%%%%%%%%%%%
\subsubsection{Frequency distribution}
%%%%%%%%%%%%%%%%%%%%%%%%%%%%%%%%%%%%%%%%%%%%%%%

\begin{figure}
\centering
\includegraphics[width=9cm]{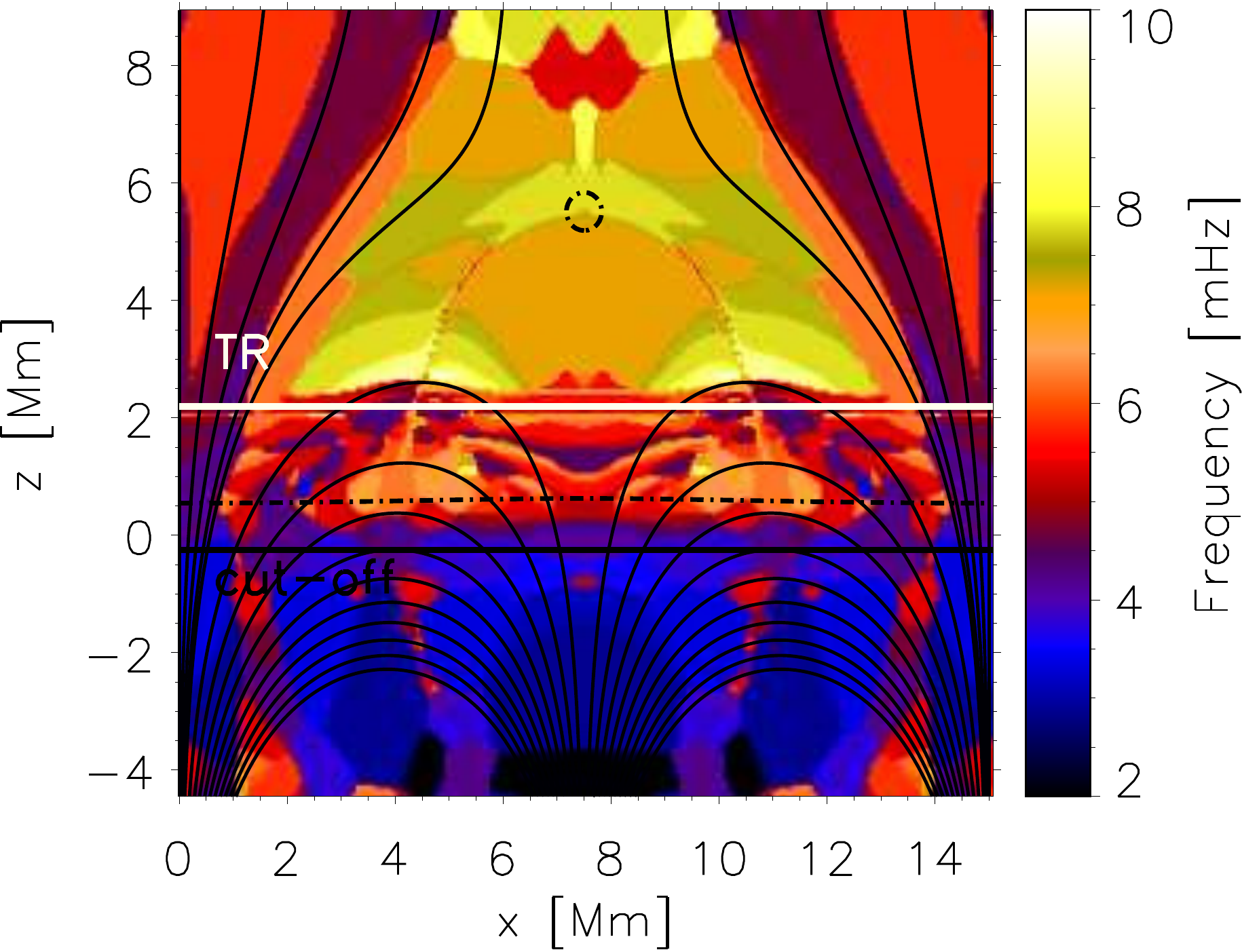}
\caption{Spatial distribution of the dominant frequencies of the vertically propagating waves for simulations with pressure pulse. Dashed-dotted black line marks $\beta=1$ contours, white solid line marks the transition region.} \label{fig:periods}
\end{figure}

The pressure pulse generates waves in a wide frequency range. The length of the simulation time series allowed us to perform the Fourier analysis of the simulations in order to determine the dominant period of the oscillations. For that, we Fourier-transformed the vertical velocity oscillations at each $(x,z)$ location of the simulation domain and calculated the frequency corresponding to the maximum power in the spectrum. These frequencies are shown in Figure \ref{fig:periods}.
%%%%%%%%%%%%%%%%%%%%%%%%%%%%%%%%%%%%%%%%%%%%%%%
%%%%%%%%%%%%%%%%%%%%%%%%%%%%%%%%%%%%%%%%%%%%%%%

Figure \ref{fig:periods} reveals that the dominant frequencies of oscillations below the photosphere lie in the 3--4 mHz range, almost independently of the magnetic structure. In particular, 3--4 mHz oscillations are the dominant ones at the photosphere, $z=0$ km. The distribution of the dominant frequencies in the chromosphere is highly dependent on the magnetic structure. Oscillations reaching the chromosphere and corona are mostly in the 5--7 mHz frequency regime, or higher. One observes that some 3--4 mHz power reaches the chromosphere at the locations of the inclined arcades. Lower frequency oscillations also reach the corona along the external part of the flux tubes where the magnetic field is more inclined. At the internal part of the flux tubes, oscillations of 6 mHz dominate. Curiously, the highest frequencies up to 10 mHz are observed above the transition region at the top parts of the inclined arcades and around the null point.

%%%%%%%%%%%%%%%%%%%%%%%%%%%%%%%%%%%%%%%%%%%%%%%
\section{Discussion and conclusions}
%%%%%%%%%%%%%%%%%%%%%%%%%%%%%%%%%%%%%%%%%%%%%%%

In this work, we have presented 2D numerical simulations for the study of the behaviour of MHD waves as they propagate from layers below the photosphere to the corona (up to 10 Mm). Our magnetic configuration (vertical flux tubes separated by an arcade-shaped structure) can be considered as representative of a quiet sun region, in terms of the weak field strength and the varying orientation. In addition, a null point exists where the magnetic field vanishes. 
\\
\\
After studying the wave behaviour at each atmospheric layer and different magnetic field inclinations, we conclude that, in a two-dimensional atmosphere, waves pass through the TR more preferably along vertical field lines. As the corona is a low-$\beta$ plasma we have mainly acoustic energy propagating into it. Part of the magnetic energy stays concentrated in the lower corona just above the arcades. The lack of magnetic energy found higher up in the corona can be explained by the large refraction that the fast magnetic waves suffer in the TR, due to the strong gradient in the Alfv\'en velocity generated by the strong density jump. 

Similar results were found by \citet{Rosenthal+etal2002}, whose work is one of the first studies of wave propagation by means of MHD simulations. These authors find that the fast magnetic waves are refracted downwards by inclined magnetic fields while those waves propagating along the field lines in almost vertical magnetic fields continue propagating upwards unaffected by the field. In a three-dimensional atmosphere, Alfv\'en waves may start to dominate the magnetic energy transport into the corona \citep[see e.g.][]{vanBallegooijen+etal2011,Khomenko+Cally2012} and the magnetic energy may become of greater importance. Non-linear effects may also influence the amount of energy transported by the waves into the corona.
\\
\\
We have also obtained that a large amount of energy propagates back downwards and upwards again in the atmosphere. On the one hand, there is the strong refraction of magnetic waves near the TR mentioned above. On the other hand, the strong gradient of the Alfv\'en velocity in the vicinity of the null point also gives rise to a strong refraction of the waves around it, capable of sending waves downwards when propagating near it. These phenomena, together with the conversion of these waves into downward acoustic waves at the $\beta=1$ layer and the turning point of these acoustic waves below the photosphere caused by the temperature gradient, lead to a continuously oscillating atmosphere, even if an instantaneous force is exerted. It is interesting to note that no periodical drivers are needed for this continuous oscillation. It must also be remarked that only the energy propagating inside the vertical flux tubes escape from the domain.  This phenomenon was also suggested by \citet{Rosenthal+etal2002}, who stated that at some moment the downward propagating waves, due to the refraction in the TR, must eventually be reflected upward again, resulting in a resonant cavity.
\\
\\
A frequency-dependent behaviour has also been obtained (see figure \ref{fig:periods}). Five-minute oscillations are channeled into the chromosphere and corona at the edges of the flux tubes where the magnetic field is more inclined, while three-minute oscillations penetrate into the higher layers in the more vertical magnetic fields. High-frequency waves can pass through the TR into the corona outside the vertical flux tubes and through the arcade. High-frequency transverse magnetic fast waves are the dominant ones in that region, even though their contribution to the magnetic energy is very small. This last result is similar to that of \citet{Fedun+etal2011b}, who also find an efficient transmission of high frequency waves into the corona inside a flux tube. Thus, it seems that waves with varying frequency may reach the corona under different conditions and magnetic field configurations.

\begin{acknowledgements}
This work is partially supported by the Spanish Ministry of Science through projects AYA2010-18029 and AYA2011-24808.
This work contributes to the deliverables identified in FP7 European Research Council grant agreement 277829, ``Magnetic connectivity through the Solar Partially Ionized Atmosphere''.
\end{acknowledgements}

%\aareferences

\end{document}